\documentclass[aps,prx,reprint,a4paper,superscriptaddress,floatfix,amsmath,amssymb,amsfonts,noshowpacs,longbibliography]{revtex4-2}
\usepackage{newtxtext,newtxmath}
\usepackage[T1]{fontenc}
\usepackage[utf8]{inputenx}
\usepackage{graphicx}
\usepackage[dvipsnames]{xcolor}
\usepackage{soul}
\usepackage[textwidth=17.5cm,textheight=23.5cm,verbose,pdftex]{geometry}
\usepackage[pdftex]{hyperref}

\hypersetup{pdfauthor={Oliver Brandt}, bookmarksnumbered=true, pdftitle={Carrier diffusion in GaN—a cathodoluminescence study. II: Ambipolar vs. exciton diffusion}, colorlinks, citecolor=blue, linkcolor=blue, urlcolor=blue}

\usepackage{floatrow}
\usepackage{csquotes
}
\DeclareMathAlphabet{\mathcal}{OMS}{cmsy}{m}{n}

\begin{document}
\title{Carrier diffusion in GaN—a cathodoluminescence study.\\ II: Ambipolar vs.\ exciton diffusion}
\author{Oliver Brandt}
\email[Electronic mail: ]{brandt@pdi-berlin.de}
\author{Vladimir M. Kaganer}
\author{Jonas Lähnemann}
\email[Electronic mail: ]{laehnemann@pdi-berlin.de}
\author{Timur Flissikowski}
\author{Carsten Pfüller}
\affiliation{Paul-Drude-Institut für Festkörperelektronik, Leibniz-Institut im Forschungsverbund Berlin e.\,V., Hausvogteiplatz 5--7, 10117 Berlin, Germany}
\author{Karl~K.~Sabelfeld}
\author{Anastasya E. Kireeva}
\affiliation{Institute of Computational Mathematics and Mathematical Geophysics, Russian Academy of Sciences, Lavrentiev Prosp.~6, 630090 Novosibirsk, Russia}
\author{Caroline Chèze}
\author{Raffaella Calarco}
\altaffiliation{Present address: Istituto per la Microelettronica e Microsistemi, Consiglio Nazionale delle Ricerche, via del Fosso del Cavaliere 100, 00133~Roma, Italy}
\author{Holger T.~Grahn}
\author{Uwe Jahn}
\affiliation{Paul-Drude-Institut für Festkörperelektronik, Leibniz-Institut im Forschungsverbund Berlin e.\,V., Hausvogteiplatz 5--7, 10117 Berlin, Germany}

\graphicspath{{./figs/}}

\begin{abstract}
    We determine the diffusion length of excess carriers in GaN by spatially resolved cathodoluminescence spectroscopy utilizing a single quantum well as carrier collector or carrier sink. Monochromatic intensity profiles across the quantum well are recorded for temperatures between 10 and 300~K. A classical diffusion model accounts for the profiles acquired between 120 and 300~K, while for temperatures lower than 120~K, a quantum capture process has to be taken into account in addition. Combining the diffusion length extracted from these profiles and the effective carrier lifetime measured by time-resolved photoluminescence experiments, we deduce the carrier diffusivity as a function of temperature. The experimental values are found to be close to theoretical ones for the ambipolar diffusivity of free carriers limited only by intrinsic phonon scattering. This agreement is shown to be fortuitous. The high diffusivity at low temperatures instead originates from an increasing participation of excitons in the diffusion process.
\end{abstract}

\maketitle

%%%%%%%%%%%%%%%%%%%%%%%%%%%%%%%%%%%%%%%%%%%%%%%%%%%%%%%%%%%%%%%%%%%%%%%%
\section{Introduction}
\label{sec:introduction}

The compound semiconductor GaN has enabled the development of light emitting diodes (LEDs) with a luminous efficacy surpassing any other light source. These devices are the building blocks of solid-state lighting, a technology whose future economic and ecological impact cannot be overstated \cite{dealmeida_2014,pust_2015}. The transformation from conventional to solid-state lighting is anticipated to result in energy savings of 130~TWh in 2020, reducing the emission of green house gases by roughly 90 million tons of CO$_{2}$ \cite{dealmeida_2014,pust_2015}. With a global market volume expected to exceed \$70 billion in 2020 for solid-state lighting alone \cite{pust_2015}, GaN is now second only to Si as the commercially most important semiconductor. However, our knowledge of the properties of this semiconductor lags far behind its commercial success.  

LEDs are bipolar semiconductor devices which are commonly modeled by the drift-diffusion equations \cite{markowich_1990}. For small-signal excitation, the material parameters entering these equations are the carrier diffusivity $D$ (with the drift mobility $\mu$ following from the Einstein relation) and the effective carrier lifetime $\tau$. The diffusion length $L = \sqrt{D \tau}$ combines these parameters in a single quantity, signifying the potential performance of the device.

For many semiconductors, systematic studies have been performed to determine these parameters and their dependence on, particularly, temperature and carrier density. As a result, the mechanisms limiting carrier diffusivity and lifetime for most technologically relevant semiconductors are fairly well understood. In particular, in several cases it has been found to be essential to take into account exciton formation \cite{zinovev_1983,hillmer_1989,erland_1993,schaefer_1996,brandt_1998,bley_1998,noltemeyer_2012,bieker_2015a,bieker_2015b,morimoto_2015,scajev_2015,naka_2016}. In materials with high exciton binding energies such as diamond \cite{morimoto_2015,scajev_2015,naka_2016}, exciton diffusion may profoundly modify the carrier diffusivity even at elevated temperatures and high carrier densities. Given the exciton binding energy of 26~meV in GaN, we would expect exciton diffusion to play an important role in this material as well \cite{brandt_1998}.

Despite its importance as the material enabling a momentous technological transition, we do not understand carrier diffusion and recombination in GaN sufficiently well to answer the question regarding the actual diffusing species: free carriers or excitons. While a variety of experimental techniques was used to investigate carrier diffusion in GaN \cite{duboz_1997,bandic_2000,chernyak_2001,cherns_2001,yakimov_2002,karpov_2003,nakaji_2005,pauc_2006a,pauc_2006b,ino_2008,lin_2009,aleksiejunas_2009,hafiz_2015,polyakov_2016,hocker_2016,liu_2016}, the main result of these efforts are values for the carrier diffusion length ranging from 30~nm to 3~\textmu m, with the majority hovering around 100--200~nm. The scatter simply reflects that the diffusion length is not a material constant, but the accumulation near 100--200~nm is at least partly a spurious result. In fact, most of these values stem from cathodoluminescence (CL) experiments deducing the diffusion length from the CL intensity around the outcrop of threading dislocations at the GaN(0001) surface. This approach seems intuitive and straightforward, but it has been recently shown to actually be a highly complex problem. Not only was the model used to extract the values of the diffusion length inappropriate \cite{yakimov_2010,sabelfeld_2017}, the basic premise underlying this method---namely, that the contrast is solely related to carrier diffusion---is attestably false \cite{kaganer_2018,kaganer_2019}. Several, if not most, of the values for the carrier diffusion length in GaN are thus incorrect and misleading. 

In any case, values for the carrier diffusion length alone do not contribute to an improved understanding of the mechanisms governing carrier diffusion. Deeper insight was gained from work utilizing the transient grating technique \cite{lin_2009,aleksiejunas_2009,scajev_2012}, which directly yields the carrier diffusivity and lifetime. In particular, \citet{scajev_2012} determined the carrier diffusivity in GaN as a function of temperature and carrier density, and compared their experimental data to calculations including the major intrinsic (phonon-mediated) scattering mechanisms. The diffusivity was found to closely follow the behavior expected for pure free-carrier diffusion down to a temperature of 80~K (the lowest in their study), but the authors did not discuss this unexpected lack of evidence for excitonic effects at this comparatively low temperature. Very recently, \citet{netzel_2020} estimated values for the carrier diffusivity in GaN between 10 and 300~K from the photoluminescence intensity of two  samples with an (In,Ga)N/GaN quantum well buried at different depth. The error bars of this estimate are considerable, but the values are systematically larger than those determined by \citet{scajev_2012}. No attempt was made to explain this difference in the absolute values quantitatively, but the authors stated that they believe the diffusivity in GaN to be dominated by excitons up to room temperature.

In the present work, we are interested in the carrier  diffusivity in GaN for temperatures between 10 and 300~K. For this purpose, we utilize CL spectroscopy in conjunction with  time-resolved photoluminescence (TRPL) experiments. This combination of techniques offers the advantage to be easily applicable to semiconductor nanostructures \cite{yoo_2008,gustafsson_2010,bolinsson_2011,yoo_2012,naureen_2013,nogues_2014}, unlike, for example, the transient grating technique. CL spectrocopy also offers, in principle, a very high spatial resolution, limited only by the generation volume. In this respect, the present work builds upon its companion paper (Ref.~\onlinecite{jahn_2020}, hereafter referred to as CD1), in which the generation volume in GaN has been experimentally determined for temperatures between 10 and 300~K and beam energies from 3 to 10~kV. We will use this generation volume once more in the subsequent paper CD3 \cite{lahnemann_2020}, in which we extract the diffusion length from the CL energy variation around threading dislocations in the same sample as used in the present work.

%%%%%%%%%%%%%%%%%%%%%%%%%%%%%%%%%%%%%%%%%%%%%%%%%%%%%%%%%%%%%%%%%%%%%%%%
\begin{figure}
\includegraphics[width=0.8\columnwidth]{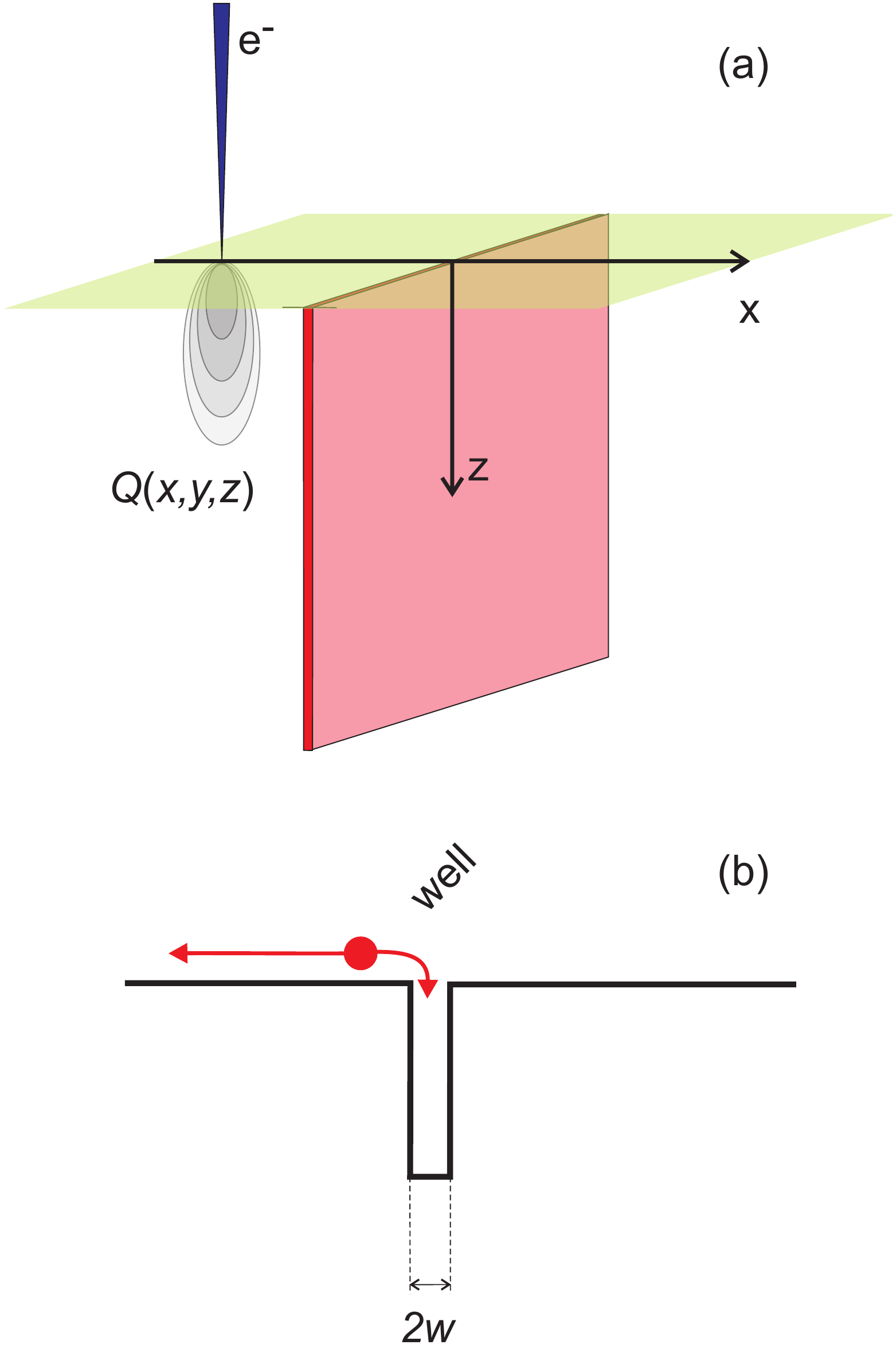}
	\caption{(a) Configuration of the CL experiment with the electron beam scanning across the well of width $2w$. (b) Sketch of the corresponding conduction band profile (not showing the piezoelectric field in the QW).}
\label{fig1}
\end{figure}
%%%%%%%%%%%%%%%%%%%%%%%%%%%%%%%%%%%%%%%%%%%%%%%%%%%%%%%%%%%%%%%%%%%%%%%%%

As the detector for diffusion, we utilize a quantum well (QW) acting as radiative sink for carriers as originally proposed for direct time-of-flight measurements by \citet{Hillmer_1986}, first utilized for CL linescans by \citet{araujo_1994a}, and previously employed for GaN in Refs.~\cite{hafiz_2015} and \cite{hocker_2016}. The single QW embedded in GaN is nominally identical to the one used in CD1, except for the absence of the thin (Al,Ga)N barriers, which were preventing diffusing carriers to reach the QW in this previous work. CL intensity profiles acquired across this QW are analyzed with a classical diffusion model taking into account quantum capture at low temperatures. The carrier diffusivity is deduced by combining these data with TRPL experiments. For temperatures below 80~K, the diffusivity reaches high values that would be expected for the ambipolar diffusion of free carriers only in the absence of any extrinsic scattering mechanism. We show that these values can be understood by considering the coexistence of thermalized populations of free carriers and excitons.

\section{Preliminary considerations} 
\label{sec:basics}

The configuration of our experiment is displayed schematically in Fig.~\ref{fig1}(a).  The focused electron beam impinges onto the cross-section of the sample along the \emph{z} direction. The beam and thus the carrier generation volume $Q(x,y,z)$ (source) is scanned across the QW along the $x$ axis normal to the well plane, with the center of the well being situated at $x=0$. During the scan, the intensity of the CL from the QW and the GaN matrix is monitored as a function of the beam position $x$. Figure~\ref{fig1}(b) depicts a scheme of the conduction band with the single QW of width 2$w$. 

The sketch ignores the presence of electrostatic fields in the structure arising from both piezoelectric and spontaneous polarization, as well as band bending resulting from charge transfer due to a finite background doping. In investigations of GaN/AlN nanowires with thin barriers, asymmetric CL intensity profiles were observed and attributed to electric fields of opposite direction at the upper and lower interfaces of the GaN quantum disk \cite{Tizei_2014,Deitz_2018,Sheng_2020}. For the present case of a single QW embedded in thick barriers, the polarization fields are concentrated in the QW, where they do not affect carrier transport, but only recombination. Selfconsistent Schrödinger-Poisson calculations of the static band profiles (not shown here) show that the weak fields remaining in the barriers are screened by the background doping. At the same time, charge transfer from donors to the well leads in turn to a symmetric band bending in its immediate vicinity, directed such as to repel electrons. However, this band bending ist not sustained under excitation of an excess carrier density comparable in magnitude to the background doping level. As a result, diffusion of carriers will dominate over their drift, and excitons will not be affected by residual weak electric fields at all because of their charge neutrality.

\begin{figure}
\includegraphics[width=0.8\columnwidth]{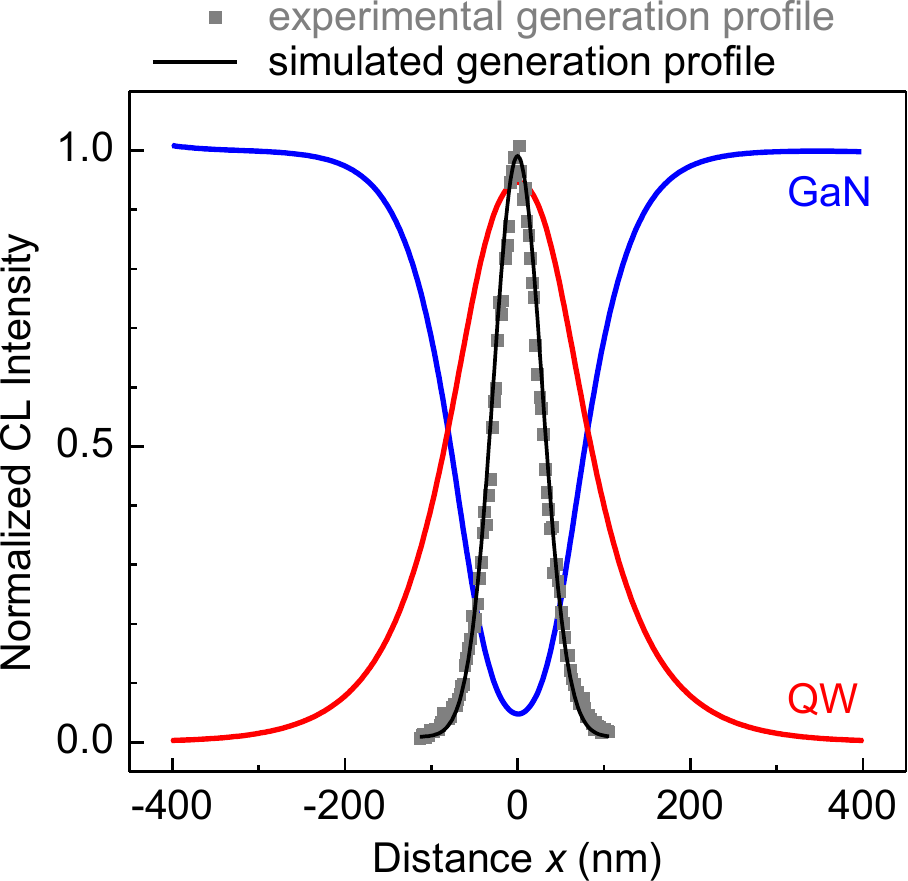}
	\caption{Exemplary carrier generation and CL intensity profiles. The squares show the experimental generation profile measured in CD1 for $V=3$~kV and $T=120$~K. The thin solid line is a fit of these data obtained by a convolution of the one-dimensional energy loss distribution $\mathcal{F}_{0}(x)$ for $V=3$~kV with a Gaussian representing the temperature-dependent broadening $\sigma =24$~nm at $T=120$~K. The solid lines show the simulated intensity profiles of the CL from the GaN matrix and the (In,Ga)N QW situated at $x=0$. These profiles are obtained by Eqs.~(\ref{eq:11}) and (\ref{eq:12}) assuming a diffusion length of $L = 100$~nm.}
\label{fig2}
\end{figure}

Without any barriers between the QW and the matrix, the QW captures carriers generated by the electron beam and diffusing in the matrix toward the QW. Carrier diffusion thus leads to an increase (decrease) of the CL intensity measured from the QW (matrix) when the electron beam is scanned across the QW along $x$. Quantitatively, the resulting CL intensity profiles across the QW are described by a one-dimensional diffusion model as detailed in Appendix \ref{appendix:diffusion}, taking into account the lateral extent of the carrier generation volume as determined in CD1 \cite{jahn_2020}. An example for the expected intensity profiles is shown in Fig.~\ref{fig2} for an acceleration voltage $V$ of 3~kV and a temperature $T$ of 120~K. The CL intensity profiles obtained by Eq.~(\ref{eq:11}) for the (In,Ga)N QW and Eq.~(\ref{eq:12}) for the GaN matrix are complementary.

%%%%%%%%%%%%%%%%%%%%%%%%%%%%%%%%%%%%%%%%%%%%%%%%%%%%%%%%%%%%%%%%%%%%%%%%%
%%%%%%%%%%%%%%%%%%%%%%%%%%%%%%%%%%%%%%%%%%%%%%%%%%%%%%%%%%%%%%%%%%%%%%%%

\section{Experimental methods} \label{sec:experiment}

For the present experiments, we utilize a 3-nm-thick In$_{0.16}$Ga$_{0.84}$N single QW embedded in the center of a 1.3-\textmu m-thick GaN layer. This sample closely resembles the one used in the preceeding study (CD1 \cite{jahn_2020}) for the experimental determination of the CL generation volume, except that it lacks the additional (Al,Ga)N barriers, and thus facilitates the capture of carriers diffusing in the GaN matrix toward the QW. The sample was fabricated by plasma-assisted molecular beam epitaxy (PAMBE) under nominally identical growth conditions as the previous one on top of a GaN(0001) template, which in turn was prepared by metal-organic chemical vapor deposition (MOCVD) on an Al$_2$O$_3$(0001) substrate. All GaN layers synthesized in this PAMBE system are \emph{n}-type due to the unintentional incoporation of O with a concentration on the order of $10^{16}$~cm$^{-3}$.

%%%%%%%%%%%%%%%%%%%%%%%%%%%%%%%%%%%%%%%%%%%%%%%%%%%%%%%%%%%%%%%%%%%%%%%%%
\begin{figure*}
\includegraphics[width=1\textwidth]{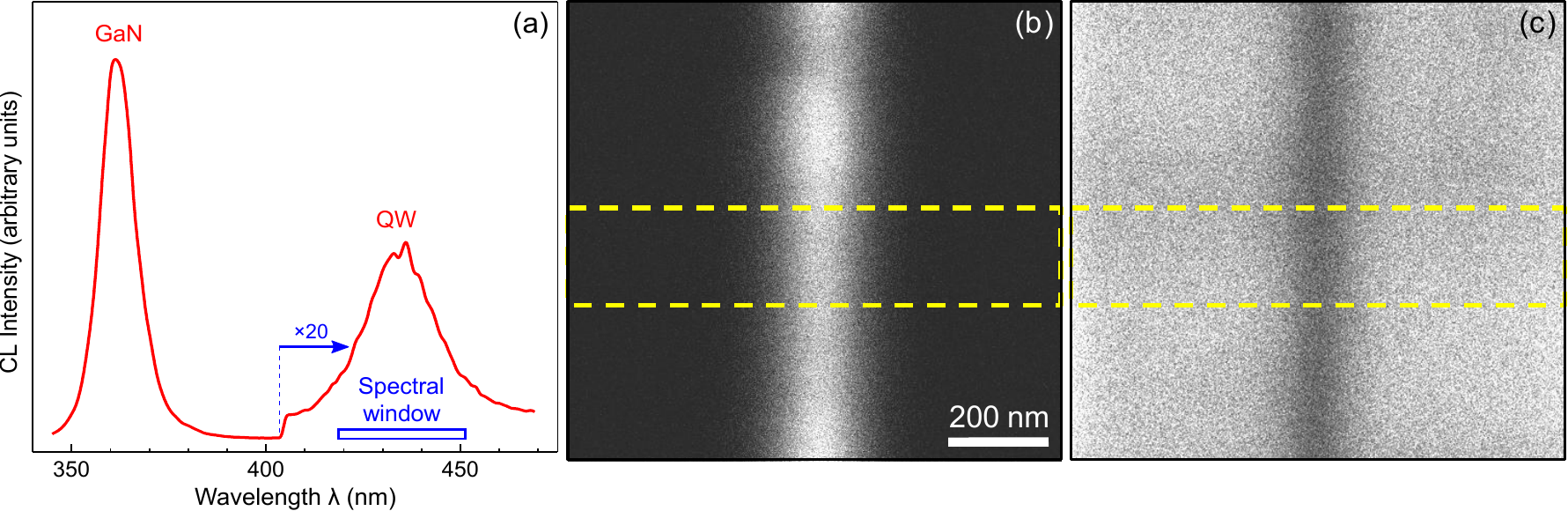}
	\caption{(a) CL spectrum acquired at the cross-section of the sample under investigation for $V=5$~kV and $T=300$~K. The spectral window used for recording monochromatic CL photon counting maps is indicated. (b) and (c) Monochromatic photon counting maps ($1.0 \times 0.9$~\textmu m$^2$) of the QW and GaN CL, respectively, at $V=5$~kV and $T=300$~K. The dashed rectangles indicate the windows for the integration of the photon counts resulting in the experimental CL profiles analyzed in this work.}
	\label{fig3}
\end{figure*}
%%%%%%%%%%%%%%%%%%%%%%%%%%%%%%%%%%%%%%%%%%%%%%%%%%%%%%%%%%%%%%%%%%%%%%%%%

The CL experiments were performed using a Gatan mono\-CL4 system and a He-cooling stage attached to a Zeiss Ultra55 SEM with a field emission gun. Cross-sectional specimens for performing CL linescans across the quantum well were obtained by cleaving the sample in air and introducing the freshly cleaved piece immediately into the SEM vacuum chamber. The acceleration voltage of the incident primary electrons was chosen to be 3 or 5~kV. The beam current was varied between 0.04 and 0.15~nA for temperatures between 10 and 300~K, respectively. A crude estimate based on the acceleration voltage, the current density, an approximation of the generation volume by a cylinder of appropriate dimensions \cite{jahn_2020}, and the carrier lifetime obtained by time-resolved photoluminescence experiments (see below) yields a cathodogenerated carrier density on the order of $10^{16}$~cm$^{-3}$ for these experiments.

Figure~\ref{fig3}(a) shows a CL spectrum of our sample integrated over an area of 1~\textmu m$^2$ at the cross-section of the sample for $T=300$~K and $V=5$~kV. The spectrum shows two CL lines centered at 362 and 435~nm originating from the GaN matrix and the In$_{0.16}$Ga$_{0.84}$N QW, respectively. CL intensity profiles crossing the QW at the cleaved edge were obtained for the respective wavelength from monochromatic CL photon counting maps such as shown in Figs.~\ref{fig3}(b) and \ref{fig3}(c), integrating over a 200-nm-wide stripe across the QW.

In order to deduce the carrier diffusivity from the measured diffusion length, the carrier lifetime was measured by time-resolved photoluminescence spectroscopy using a synchroscan streak camera connected to a spectrometer with 3~meV spectral resolution.  The top surface of the sample was excited by the second harmonic of 200~fs laser pulses obtained from an optical parametric oscillator pumped by a Ti:sapphire fs laser system.  The excitation energy was 3.814~eV with an energy fluence per pulse of about 0.1~$\mu$J\,cm$^{-2}$, corresponding to a photogenerated carrier density of $10^{16}$~cm$^{-3}$. The sample temperature was controlled by a constant flow cryostat between 10 and 300~K.

\section{Results and Discussion}\label{sec:results}

\subsection{Modelling of CL profiles}
\label{sec:modelling}

\begin{figure*}
	\floatbox[{\capbeside\thisfloatsetup{capbesideposition={right,center},capbesidewidth=4.5cm}}]{figure}[\FBwidth]
	{\caption{Comparison of experimental CL intensity profiles extracted from photon counting maps recorded at 3~kV  (symbols) with their fits (solid lines) by the diffusion model described in Appendix \ref{appendix:diffusion}. The dashed lines show the corresponding CL generation profile. At 300~K, the model provides an excellent fit to the profiles from both the GaN matrix (a) and the QW (c). In contrast, neither of the profiles obtained at 20~K are adequately described by the diffusion model, as seen in panels (b) and (d). In particular, the top-hat shape of the QW CL profile (d) is incompatible with purely diffusive carrier transport.}
	\label{fig4}}
	{\includegraphics[width=0.7\textwidth]{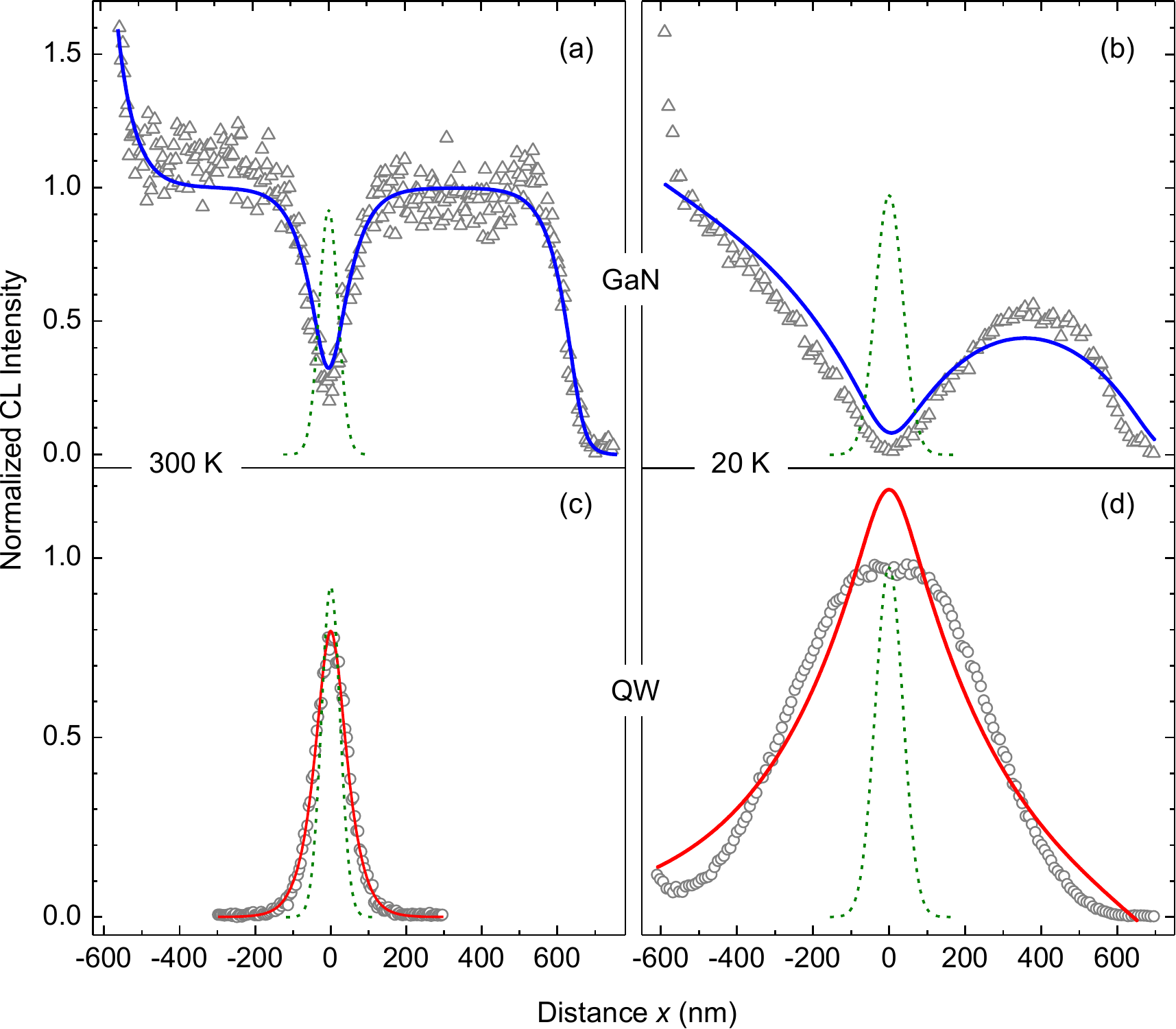}}
\end{figure*}

Figure \ref{fig4} displays a comparison of experimental CL intensity profiles extracted from photon counting maps recorded at 3~kV (symbols) and their fits (lines) by the diffusion model described in Appendix \ref{appendix:diffusion}. The profiles obtained at 300~K [Figs.~\ref{fig4}(a) and \ref{fig4}(c)] are reproduced fairly well by the model. The CL intensity profile from the GaN matrix in Fig.~\ref{fig4}(a) spans the entire MBE-grown GaN/(In,Ga)N/GaN structure from its interface with the MOCVD template at about $-650$~nm to the sample surface at about 650~nm. Because of the higher luminous efficiency of the MOCVD template as compared to the PAMBE-grown GaN layer on top, carrier diffusion toward this interface results in a higher CL intensity. Analogously, the intensity decrease close to the surface results from carriers diffusing toward the surface and recombining there nonradiatively. The widths of these transition regions and the width of the central minimum due to the quantum well (as well as the maximum in the complementary CL intensity profile from the QW) is thus expected to be determined by the diffusion length. This interpretation is supported by the fact that a single value for the diffusion length of $(40 \pm 5)$~nm reproduces the entire profile in Fig.~\ref{fig4}(a) as well as the complementary profile of the QW CL in Fig.~\ref{fig4}(c). The error of the latter fit is significantly lower compared to the former, since the QW CL profile is neither affected by inhomogeneities in the matrix, nor by the additional boundaries at the bottom interface and top surface.

We obtain fits of similar quality for temperatures down to 120~K, but at lower temperatures, the shape of the profiles progressively deviates from the one expected for purely diffusive transport. As an example, Figs.~\ref{fig4}(b) and \ref{fig4}(d) display the CL intensity profiles obtained at 20~K from the GaN matrix and the QW, respectively. In both cases, the diffusion model is clearly not able to reproduce the experimental lineshape. In particular, it proves to be impossible to obtain a fit of both the central minimum and the bottom and top boundaries in the GaN CL profile with a single value for the diffusion length. The fit shown represents a compromise and returns a value that is too large for the boundaries, but too small for the central minimum due to the QW. Moreover, at the position of the QW, the intensity of the GaN CL is essentially zero, and the QW profile exhibits a pronounced top-hat shape. These features are incompatible with a purely diffusive process, and signify that at low temperatures, the QW captures carriers in its vicinity more efficiently than possible by diffusion.

%%%%%%%%%%%%%%%%%%%%%%%%%%%%%%%%%%%%%%%%%%%%%%%%%%%%%%%%%%%%%%%%%%%%%%%%%
%%%%%%%%%%%%%%%%%%%%%%%%%%%%%%%%%%%%%%%%%%%%%%%%%%%%%%%%%%%%%%%%%%%%%%%%

%QUANTUM CAPTURE

Indeed, such a capture process is known to be inherent to QWs since the pioneering work of \citet{brum_1986}. In later work, the phenomenon has become known as \enquote{quantum capture}, a term representing the resonant capture of hot carriers relaxing directly into virtual bound states of the QW close to the continuum edge by LO phonon emission. Early studies, focused on the group-III arsenides, have established the physical principles of this process and estimated its efficiency \cite{brum_1986,polland_1988,fujiwara_1992,kan_1992,blom_1993,vassilovski_1995}. More recently, the group-III nitrides moved into the focus of interest \cite{mansour_1995,stavrou_1998,zakhleniuk_1999,fan_2004,stavrou_2011,vallone_2015,vallone_2017}. In both cases, it has been shown that quantum capture times may be significantly shorter than the momentum relaxation time at low temperature, which is a prerequisite for a process relying on the spatial coherence of quantum states. The capture process thus occurs quasi-ballistically, but unlike classical ballistic transport, quantum capture is directional, and hence dominates over all other transport mechanisms from a certain (temperature-dependent) distance from the QW.

\begin{figure*}
	\includegraphics[width=1\textwidth]{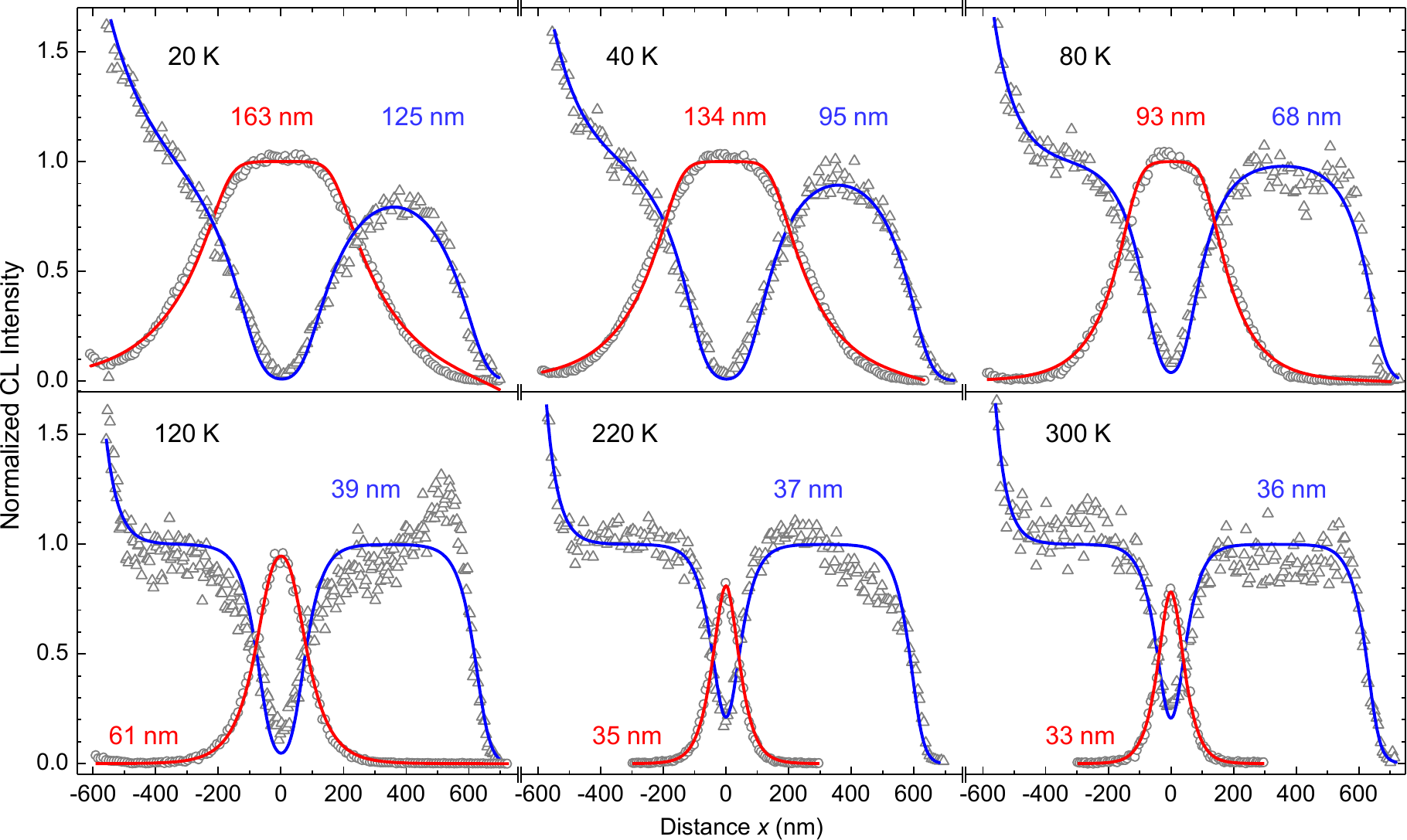}
		\caption{Comparison of experimental CL intensity profiles recorded for the GaN matrix (triangles) and the QW (circles) at 3~kV and temperatures between 20 and 300~K with their fits (lines) by the extended diffusion model described in Appendix \ref{appendix:diffusion}, taking into account both diffusion and quantum capture. The values for the diffusion length returned from the fits are indicated next to the respective profiles. Note that the intensity quenching close to the sample edge at the right is partly due a scattering of electrons out of the sample when the electron beam approaches the edge due to the vicinity of the free top surface (see also Appendix \ref{appendix:diffusion}).}
		\label{fig5}
	\end{figure*}

Figure~\ref{fig5} displays a comparison of experimental (symbols) and simulated (lines) QW- and GaN-CL profiles acquired at temperatures between 20 and 300~K and for an acceleration voltage of 3~kV. The simulated CL profiles are obtained by considering one additional free parameter, the quantum capture length $L_{Q}$, which we approximate simply by an effective width of the QW in the diffusion model used already above and  discussed in Appendix \ref{appendix:diffusion}. This approximation may appear crude since it corresponds to an instantaneous process, but it contains the most important feature of quantum capture: a drastically enhanced capture rate for carriers crossing a certain critical threshold distance from the QW.

Evidently, the fits of the experimental profiles at low temperatures are much improved compared to the pure diffusion model used for the examples shown in Figs.~\ref{fig4}(b) and \ref{fig4}(d). In particular, the simulations now reproduce the top-hat profiles for the QW observed below 80~K and the deep minimum for the GaN matrix (Fig.~\ref{fig5}, upper row). The values for the diffusion length obtained from the QW- and the GaN-CL profile agree perfectly between 220 and 300~K, but start to deviate at lower temperatures. A closer look at the GaN-CL profiles reveals that the fit tends to underestimate the diffusion length in the central QW-induced minimum to match the slope of the extrema at the outer boundaries. For temperatures below 80~K, these regions additionally start to overlap due to the increasing diffusion length.

We have performed several measurements such as depicted in Figs.~\ref{fig3}(b) and \ref{fig3}(c), and estimated the diffusion length such as shown in Fig.~\ref{fig5}. The values extracted from the profiles of the GaN-CL and the QW-CL signal follow the same trend, but the former exhibit a larger scatter, which we believe to be due to the influence of the bottom and top boundaries as discussed above. In the following, we thus focus on values for the diffusion length obtained from the QW-CL profiles.

\subsection{Temperature dependence of $\boldsymbol{L}$ and $\boldsymbol{D}$: minority carrier and ambipolar diffusion}
\label{sec:temperature}

\begin{figure}
    \includegraphics[width=0.85\columnwidth]{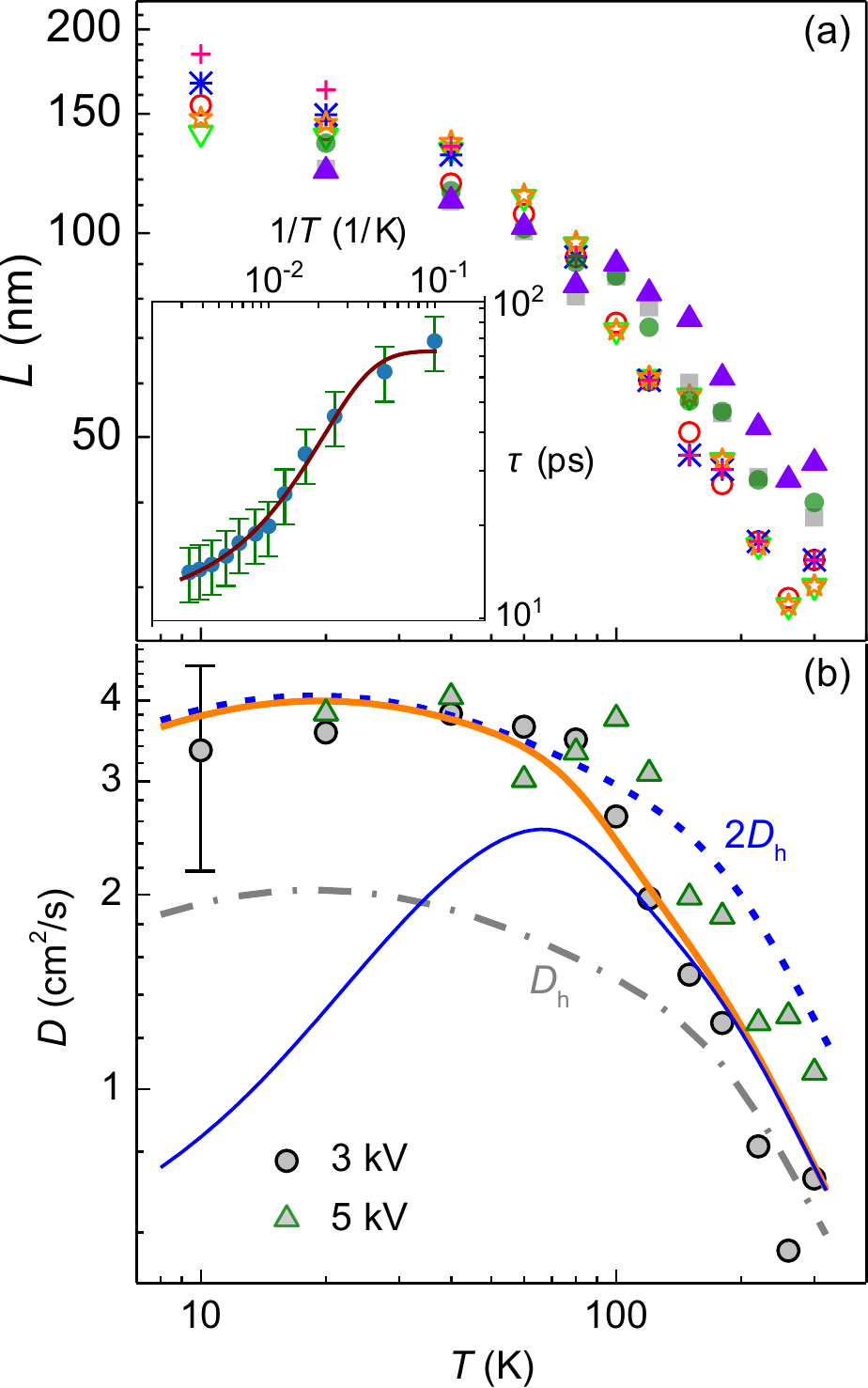}
	\caption{(a) Diffusion length $L$ derived from intensity profiles of the QW CL as a function of $T$ between 10 and 300~K. The different symbols represent 8 different measurements using acceleration voltages of either 3 or 5~kV. The inset shows a double-logarithmic plot of the carrier lifetime $\tau$ vs.\ inverse temperature $1/T$. The symbols represent average values, and the error bars indicate the typical variation for different locations on the sample. The line represents an Arrhenius fit with a single activation energy. (b) Comparison of experimental (symbols) and theoretical (lines) carrier diffusivity $D$. Averaged data acquired at 3 and 5~kV are represented by circles and triangles, respectively. The error bar of $D$ resulting from the uncertainty in $L$ and $\tau$ is shown exemplary for the data point at 10~K. The dash-dotted and dashed lines show the theoretical limits of the minority carrier (hole) and maximum ambipolar diffusivity, respectively, set by phonon scattering. The thick solid line represents a fit of the data with Eq.~(\ref{eq:ambipolar}) taking into account the temperature dependence of both the background and excess carrier density. The thin solid line shows a fit considering ionized impurity scattering in addition to phonon scattering.}
	\label{fig6}
\end{figure}

Figure~\ref{fig6}(a) shows the results of 8 different measurements performed at both 3 and 5~kV and temperatures between 10 and 300~K. The data consistently show an increase of the diffusion length by a factor of 5 from 300 to 10~K. This finding is not overly surprising, since both the diffusivity and the carrier lifetime tend to increase with decreasing temperature, and so should the diffusion length. Hence, the values of the diffusion length alone as shown in Fig.~\ref{fig6}(a) do not contribute to an understanding of the diffusion process. We can neither pinpoint the mechanism limiting the diffusion length, nor can we deduce the nature of the diffusing species. 
%Worse, we cannot even begin to understand whether or not the values measured are physically sound or not. This point is essential, as demonstrated most clearly by the \enquote{diffusion length} extracted from CL linescans across dislocations (which are actually largely unaffected by diffusion \cite{kaganer_2018,kaganer_2019}). 
To go further, we thus need to determine the carrier diffusivity instead of the diffusion length.

To obtain the diffusivity as a function of temperature, we measure the effective carrier lifetime $\tau$ in the same temperature range by time-resolved photoluminescence experiments exciting the top GaN(0001) layer (for exemplary intensity transients and the extraction of $\tau$, see Appendix \ref{appendix:lifetime}). The average lifetimes obtained as well as the typical variation observed in different locations on the sample are shown in the inset of Fig.~\ref{fig6}(a). With values ranging from 15~ps at 300~K to 80~ps at 10~K, they are about a factor of 2--3 shorter than those observed for the MOCVD template measured under the same (small-signal) conditions. These values thus represent bulk nonradiative lifetimes, and are neither governed by surface recombination nor by threading dislocations (as shown in CD3, the dislocation density in the PAMBE layers is identical to that of the template, and amounts to $5 \times 10^8$~cm$^{-2}$) \cite{lahnemann_2020}. The Arrhenius fit of the data returns an activation energy of $(7.6 \pm 1.9)$~meV, suggesting that the reduction of the lifetime is controlled by the thermal dissociation of donor-bound excitons, having a binding energy of 7~meV. 

%%%%%%%%%%%%%%%%%%%%%%%%%%%%%%%%%%%%%%%%%%%%%%%%%%%%%%%%%%%%%%%%%%%%%%%%%
%%%%%%%%%%%%%%%%%%%%%%%%%%%%%%%%%%%%%%%%%%%%%%%%%%%%%%%%%%%%%%%%%%%%%%%%%

Figure~\ref{fig6}(b) shows the experimental values of the diffusivity $D$ obtained by combining the diffusion length $L$ with the carrier lifetime according to $D = L^2/\tau$. For clarity, we have averaged the data sets shown in Fig.~\ref{fig6}(a) and now distinguish only between measurements performed at 3 (circles) and 5~kV (triangles). Likewise, $\tau$ is an average of three sets of measurements performed on different locations of the sample. The diffusivity is seen to steeply increase from about 0.8 to 3.5~cm$^2$/s when lowering the temperature from 300 to 80~K, and then to more or less saturate at this value. These values agree very well with those determined by \citet{scajev_2012} in this temperature range, which the authors interpreted as representing minority carrier or ambipolar diffusion, depending on excitation density.

Indeed, the ambipolar diffusivity \cite{vanroosbroeck_1953}
\begin{equation}
	D_{a}=\frac{(n_0+\Delta n+\Delta p) D_e D_h}{(n_0+\Delta n) D_e+\Delta p D_h}
	\label{eq:ambipolar}
\end{equation}
with the diffusivities of electrons and holes $D_{e}$ and $D_{h}$, respectively, varies between $D_{h}$ for $n_{0} \gg \Delta n$ and $2 D_{h}$ for $n_{0} \ll \Delta n$, where $n_0$ is the background doping density, and $\Delta n \approx \Delta p$ is the excess carrier density generated by the electron beam. To calculate $D_{e}$ and $D_{h}$, we take into account the major intrinsic scattering mechanims for electrons and holes caused by acoustic [Eqs.~(\ref{eq:piezo}) and (\ref{eq:def})] as well as by polar optical [Eq.~(\ref{eq:polar})] phonons. Regarding acoustic phonons, we take into account both deformation potential [Eq.~(\ref{eq:def})] and piezoelectric [Eq.~(\ref{eq:piezo})] scattering (for details of this calculation as well as a compilation of the various material parameters entering it, see Appendix \ref{appendix:mobility}). Note that the latter was neglected by \citet{scajev_2012}, but becomes important for temperatures below 100~K. Our value for the hole diffusivity at 100~K is thus slightly smaller than reported in Ref.~\onlinecite{scajev_2012} (1.48 vs.\ 2~cm$^2$/s), but in almost exact agreement with the value (1.46~cm$^2$/s) obtained within an \emph{ab initio} Boltzmann transport approach \cite{ponce_2019}. The dash-dotted and dashed lines in Fig.~\ref{fig6}(b) show the theoretical dependence of the minimum and maximum ambipolar diffusivity on temperature, respectively. Clearly, our experimental results are situated just in between these limits, suggesting that carrier diffusion is limited by holes at high temperatures, and by full ambipolar transport at low temperatures.

%%%%%%%%%%%%%%%%%%%%%%%%%%%%%%%%%%%%%%%%%%%%%%%%%%%%%%%%%%%%%%%%%%%%%%%%%
%%%%%%%%%%%%%%%%%%%%%%%%%%%%%%%%%%%%%%%%%%%%%%%%%%%%%%%%%%%%%%%%%%%%%%%%%

The transition between these two regimes is largely a consequence of the temperature dependence of both $n_0$ and $\Delta n$ when the temperature is lowered from 300 to 10~K. The former is given by the familiar expression \cite{seeger_1989}

\begin{equation}
	\frac{n_0 (n_0 +N_A)}{N_D-N_A-n_0}=\frac{N_c}{2} e ^{-E_D/ k_B T}
	\label{eq:carrconc}
\end{equation}
with the donor and acceptor densities $N_{D}$ and $N_{A}$, the donor ionization energy $E_D$, and the effective conduction-band density of states

\begin{equation}
	N_{c}=\frac{1}{\sqrt{2}} \left(\frac{m_e  k_B T}{\pi \hbar ^2}\right)^{3/2},
	\label{eq:effdos}
\end{equation}
where $m_e$ is the density-of-states mass of the electron, and the other symbols have their usual meaning. The excess carrier density is given by $\Delta n \approx \Delta p  = G_e \tau(T)$ with the generation rate $G_e$ and the carrier lifetime $\tau (T)$ obtained from the time-resolved photoluminescence measurements.

Secondary ion mass spectrometry (SIMS) performed on GaN test structures grown in the same MBE system as the present sample reveals O and C concentrations of about $4 \times 10^{16}$ and $2 \times 10^{16}$~cm$^{-3}$, which we thus take here as the values for $N_D$ and $N_A$, respectively. The ionization energy of O is taken to be $E_{D} = 30$~meV \cite{wysmolek_2002}, resulting in an electron background density of $1.8 \times 10^{16}$~cm$^{-3}$ at room temperature. This value is close to those obtained by capacitance-voltage profiling on samples from this MBE system.

The thick solid line in Fig.~\ref{fig6}(b) shows the fit of Eq.~(\ref{eq:ambipolar}) taking into account  Eqs.~(\ref{eq:piezo})--(\ref{eq:polar})  to the data points averaged over all measurements at 3~kV with the generation rate $G_{e}$ as the only free parameter. The fit mediates between the diffusivity of holes at high temperatures and the maximum ambipolar diffusivity at low temperatures for a generation rate resulting in an excess carrier density at 300~K lower than the background density ($4 \times 10^{15}$~cm$^{-3}$ vs.\ $1.8 \times 10^{16}$~cm$^{-3}$). The rate of increase in $D$ between 300 and 100~K results primarily from the rapid freeze-out of background electrons at O donors, governed by the ionization energy of the latter. Furthermore, the longer effective PL lifetime results in an excess carrier density of $3 \times 10^{16}$~cm$^{-3}$ at 10~K.

At first glance, the data seem to imply that the carrier diffusion in GaN occurs exclusively by uncorrelated electron-hole pairs. However, when we recall that an ambipolar diffusivity of 3.5~cm$^2$/s at 10~K corresponds to a hole drift mobility of about 2000~cm$^2$/Vs, it becomes clear that this interpretation is not a credible one. In fact, this value is larger than the highest \emph{electron} drift mobility measured to date for free-standing GaN with considerably higher purity than the present layer \cite{look_2001}. 

It is easy to understand why our simulations predict such a high diffusivity (or mobility): we have so far taken into account only intrinsic phonon-induced scattering mechanisms, while the low-temperature mobility of charge carriers in bulk semiconductors is invariably limited by their interaction with charged defects, such as ionized impurities \cite{seeger_1989,chattopadhyay_1981} and (in the case of epitaxial GaN layers) threading dislocations \cite{look_1999}. When taking into account ionized impurity scattering \cite{ridley_1977} with a density of ionized impurities of $N_{i} = n_0 + 2N_A$ [Eq.~(\ref{eq:ii})], we obtain the thin solid line in Fig.~\ref{fig6}(b) exhibiting the characteristic decrease of the diffusivity with decreasing temperature. This effect is still moderate in the present case because the total carrier density $n_0 + \Delta n$ remains high down to low temperatures, thus effectively screening the scattering potential of charged impurities. The decrease would also be more pronounced if dislocation scattering (with a density of $5 \times 10^8$~cm$^{-2}$) were taken into account in addition. In any case, it is clear that a realistic treatment of free carrier scattering at low temperatures will lead to a drastic deviation of the theoretically predicted and experimentally observed diffusivities, which are (in the context of free carrier diffusion) anomalously high.

\subsection{Temperature dependence of $\boldsymbol{L}$ and $\boldsymbol{D}$: exciton diffusion}
\label{sec:exciton}

The natural explanation of this apparent discrepancy is that we are not observing the diffusion of geminate electron-hole pairs at low temperatures, but of excitons. Excitons as neutral entities interact only weakly with charged defects, and are thus expected to retain a comparatively high diffusivity at low temperatures \cite{zinovev_1983,brandt_1998}. 

\begin{figure}
\includegraphics[width=0.95\columnwidth]{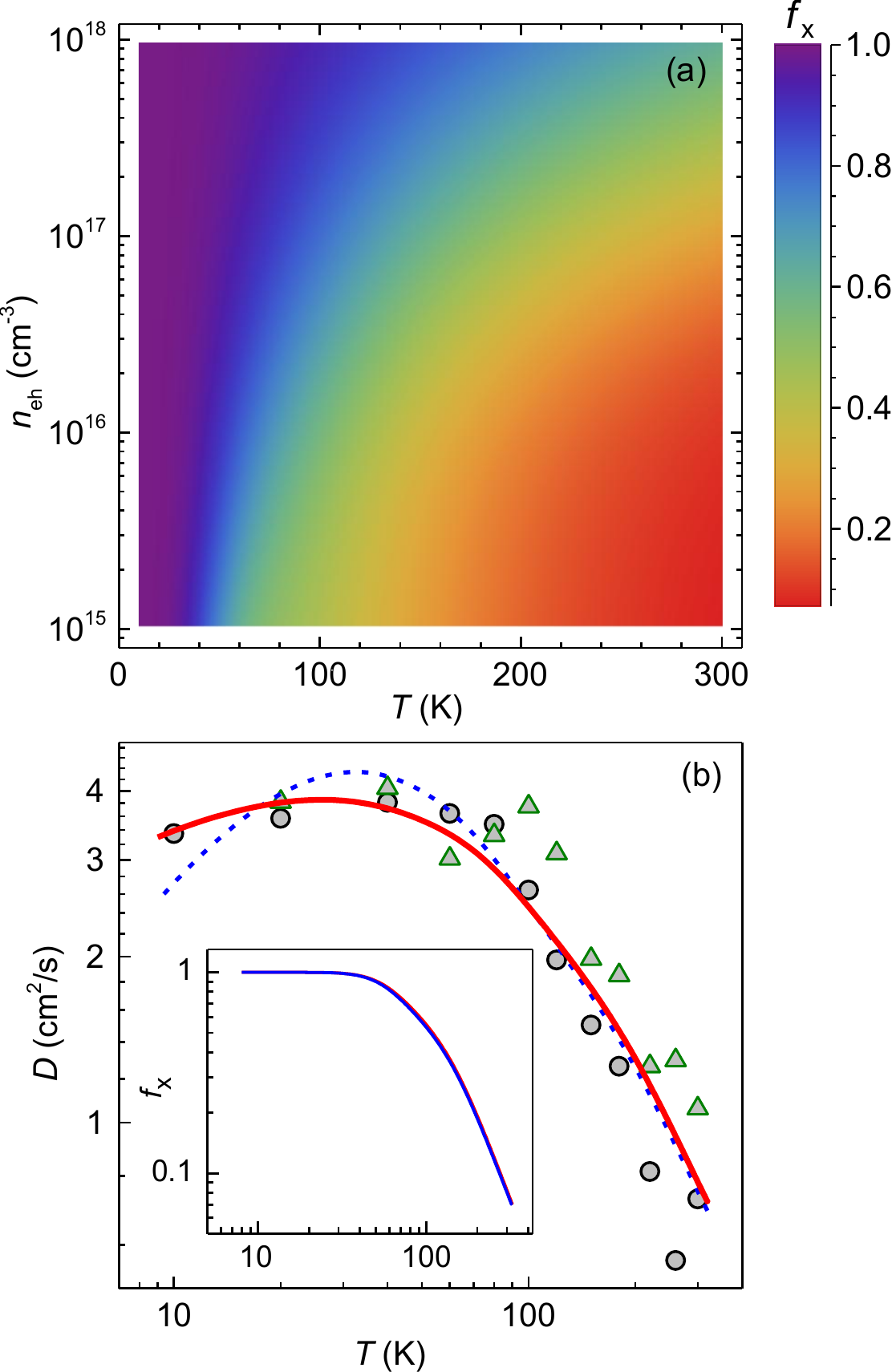}
\caption{(a) Phase diagram of the coupled exciton-carrier system in GaN. The color code represents the exciton fraction $f_x=n_x/n_\text{eh}$ according to Eq.~(\ref{eq:saha}). (b) Comparison of experimental (symbols) and theoretical (lines) carrier diffusivity $D$. Averaged data acquired at 3 and 5~kV are represented by circles and triangles, respectively. The lines show fits with Eq.~(\ref{eq:excitonic}) considering the simultaneous diffusion of free carriers and excitons and assuming either a vanishing contribution from piezoelectric (dashed) or neutral impurity scattering (solid). The inset shows the (essentially coinciding) exciton fractions $f_{x}$ returned by these fits.}
\label{fig7}
\end{figure}

Furthermore, in GaN with its exciton binding energy of 26~meV \cite{volm_1996,rodina_2001}, we also expect that the diffusing species change in this temperature range from mostly free carriers to mostly excitons. Figure~\ref{fig7}(a) displays a phase-diagram of the coupled free carrier/exciton system in GaN as predicted by the law of mass action, or Saha equation \cite{saha_1920,ebeling_1976,gourley_1982,bieker_2015b}. The diagram depicts the fraction of electron-hole pairs $f_x = n_x/n_ {\text {eh}}$ forming excitons, with the total density of cathodogenerated electron-hole pairs $n_ {\text {eh}}$. The density of unbound, free carriers is then given by $\Delta n =\Delta p = \left (1 - f_x \right) n_ {\text {eh}}$. Their relation is given by

\begin{equation}
	\frac{\left(n_ 0 + \Delta n \right)\Delta p }{n_x}= \frac{N_ {\text {cv}}} {2}e^{-E_X/k_B T}
	 \label{eq:saha}
\end{equation}
with the exciton binding energy $E_{X}$ (assumed here to be independent of $n_\text{eh}$) and the effective reduced density of states which is defined analogously to Eq.~(\ref{eq:effdos}) but with $m_e$ replaced by the reduced density-of-states mass of the exciton. Note that many-body effects reduce the exciton binding energy for carrier densities approaching the Mott density (around $2 \times 10^{18}$~cm$^{-3}$) \cite{choi_2001,kyhm_2011}, so that $f_x$ is overestimated in the high-density regime. Since the carrier density in our experiments is well below $10^{17}$~cm$^{-3}$ (see Sec.~\ref{sec:experiment}), these high density effects should thus not affect our results.

%%%%%%%%%%%%%%%%%%%%%%%%%%%%%%%%%%%%%%%%%%%%%%%%%%%%%%%%%%%%%%%%%%%%%%%%%
%%%%%%%%%%%%%%%%%%%%%%%%%%%%%%%%%%%%%%%%%%%%%%%%%%%%%%%%%%%%%%%%%%%%%%%%%%  

We can estimate the contribution of excitons to diffusion by examining the experimental data depicted in Fig.~\ref{fig6}(b). The crossover from pure hole to the maximum ambipolar diffusivity between 300 and 100~K necessitates an excess carrier density exceeding the background doping level, i.\,e., about $10^{16}$~cm$^{-3 }$ at 100~K. For this carrier density, Fig.~\ref{fig7}(a) informs us that the excitonic fraction $f_x$ is larger than 0.6 for temperatures lower than 80~K. 

Following these considerations, we calculate the total diffusivity $D_{c}$ of the coupled carrier-exciton population by \cite{scajev_2015}

\begin{equation}
	D_c=\left(\frac{f_x}{D_x}+\frac{1-f_x}{D_{a}}\right)^{-1},
	\label{eq:excitonic}
\end{equation}
where $D_{x}$ is the exciton diffusivity (see Appendix \ref{appendix:mobility}). 

Figure \ref{fig7}(b) shows our experimental data with two fits of Eq.~\ref{eq:excitonic} to the averaged data aquired at 3~kV. In addition to ionized impurity scattering for free carriers, we assume that either piezoelectric scattering or scattering with neutral impurities [Eq.~(\ref{eq:ni})] with a density $N_{n}$ is the process limiting the low-temperature diffusivity of excitons. The solid line shows a fit based on the former assumption, i.\,e., $N_{n} = 0$, and the free parameters are the generation rate $G_e$ and the electromechanical coupling coefficient $K_\text{av}$. The best fit is obtained with $K_\text{av} = 0.14$, which is larger than the one for electrons and holes, and is thus an unlikely value for excitons for which we would expect a weaker piezoelectric coupling \cite{zinovev_1983}. The dashed line is based on the latter assumption, namely, $K_\text{av} = 0$, returning $N_{n} = 3 \times 10^{17}$~cm$^{-3}$. This density is significantly higher than the one of neutral donors (which, at low temperatures, is expected to be $N_D -N_A$), but close to the typical concentration of the isoelectronic impurity B detected by SIMS in our samples. In any case, the accuracy of our data is insufficient to distinguish between these two opposite extremes, but it is clear that the data could be fit very well with sensible values for both $K_\text{av}$ and $N_{n}$.  The high diffusivities measured at low temperatures are thus a natural consequence of the progressive freeze-out of free electrons and holes into excitons with their lower sensitivity to defect-induced scattering compared to free carriers.

In this context, it is of interest to compare our results with those of recent measurements of \citet{netzel_2020}, who  deduced the temperature-dependent carrier diffusivity in MOCVD-grown GaN from the photoluminescence intensity of buried (In,Ga)N/GaN quantum wells combined with separate time-resolved photoluminescence experiments yielding the carrier lifetime.  While the experimental uncertainty of the results is considerable, all values below 200~K are systematically higher (by a factor of at least two) than ours or those of \citet{scajev_2012}. In view of the discussion above, it is tempting to ascribe these high diffusivities to enhanced exciton diffusion in a sample of higher purity. However, this interpretation does not hold up to closer inspection. For the photogenerated carrier densities ($< 10^{15}$~cm$^{-3}$) employed in these experiments, excitons become important for the total diffusivity only at temperatures below 100~K [cf.\ Fig.~\ref{fig7}(b)], and cannot be responsible for the high diffusivities observed up to 200~K. As we show in Appendix \ref{appendix:diffusion}, these high values instead result from the simplistic model used in Ref.~\cite{netzel_2020} to deduce the diffusion length. The larger the actual diffusion length is, the more this model overestimates its value, leading in turn to values of $D \propto L^2$ that are significantly too high. For a surface recombination velocity of $4 \times 10^4$~cm/s \cite{onuma_2012,bulashevich_2016}, $D$ is overestimated by a factor of about 1.3 at high and 1.9 at low temperatures. Within their error bars, the values of \citet{netzel_2020} are thus consistent with ours and those of \citet{scajev_2012}, but they cannot help to answer the question raised above about the piezoelectric scattering rate of excitons.

%%%%%%%%%%%%%%%%%%%%%%%%%%%%%%%%%%%%%%%%%%%%%%%%%%%%%%%%%%%%%%%%%%%%%%%%
\section{Summary and conclusion}
\label{sec:summary}

Using a single (In,Ga)N QW as a radiative sink for carriers, we have investigated the carrier diffusion in GaN in a temperature range from 10 to 300~K. The diffusion length has been deduced from CL intensity profiles recorded across the QW. In conjunction with independent measurements of the carrier lifetime, we have determined the carrier diffusivity, which we have found to assume unexpectedly high values at low temperatures, in particular, higher than possible for the ambipolar diffusion of free carriers in GaN. We have demonstrated that these values are due to the fact that excitons dominate the diffusion process for temperatures below 80~K. 

The exciton diffusivity at low temperatures is limited by either piezoelectric or neutral impurity scattering. To distinguish between these scenarios will require dedicated experiments with improved accuracy on samples with controlled densities of neutral scattering centers. Experimental methods suitable for such a study would be the transient grating technique on plain GaN layers or time-resolved CL spectroscopy on custom-designed single QW samples as used in the present study. Both of these techniques have the decisive advantage of a simultaneous measurement of carrier diffusivity and lifetime. Only the latter technique is applicable to three-dimensional nanostructures such as nanowires and can thus be employed to answer the numerous unresolved questions regarding the carrier transport in such structures. 

\begin{acknowledgments}
The authors are indebted to Achim Trampert for a critical reading of the manuscript and stimulating discussions. Special thanks are due to Lutz Geelhaar and Henning Riechert for their continuous encouragement and support. K.~K.~S.\ and A.~E.~K.\ acknowledge funding from the Russian Science Foundation under grant N 19-11-00019.
\end{acknowledgments}

\appendix

\section{Solution of the carrier diffusion problem}
\label{appendix:diffusion}

\subsubsection*{Formulation of the diffusion problem}

The three-dimensional problem of diffusion of the electron-hole pairs produced by the source $Q(x,y,z)$ sketched in Fig.~\ref{fig1}(a) can be reduced to a one-dimensional problem in the same way as it has been done in CD1. The formalism developed here applies equally to minority-carrier, ambipolar, and exciton diffusion, but for simplicity we exclusively refer to excitons in all what follows.

The top surface $z=0$ of the sample, which is the \textit{M}-plane of GaN, possesses a low surface recombination velocity \cite{corfdir_2014}. Hence, the diffusing excitons are reflected from this surface, rather than being absorbed at it. The trajectory of the diffusing exciton after its reflection can be mirrored to the other half-space and the exciton path can be considered as a diffusion process in the whole space. The (In,Ga)N well is also extended to the other half-space. A more formal treatment of the extension of the diffusion problem from half- to whole-space, in the case of the reflection boundary condition at the planar surface, has been discussed in Ref.~\cite{sabelfeld_2017}. We therefore consider a one-dimensional source 
\begin{equation}
\bar{Q}(x)=\intop_{-\infty}^{\infty}dy\intop_{0}^{\infty}dz\,Q(x,y,z).\label{eq:1}
\end{equation}
The distribution $\bar{Q}(x)$ has been experimentally determined in CD1 for different temperatures and acceleration voltages, and we use these in the present work.

\begin{figure}
\includegraphics[width=1\columnwidth]{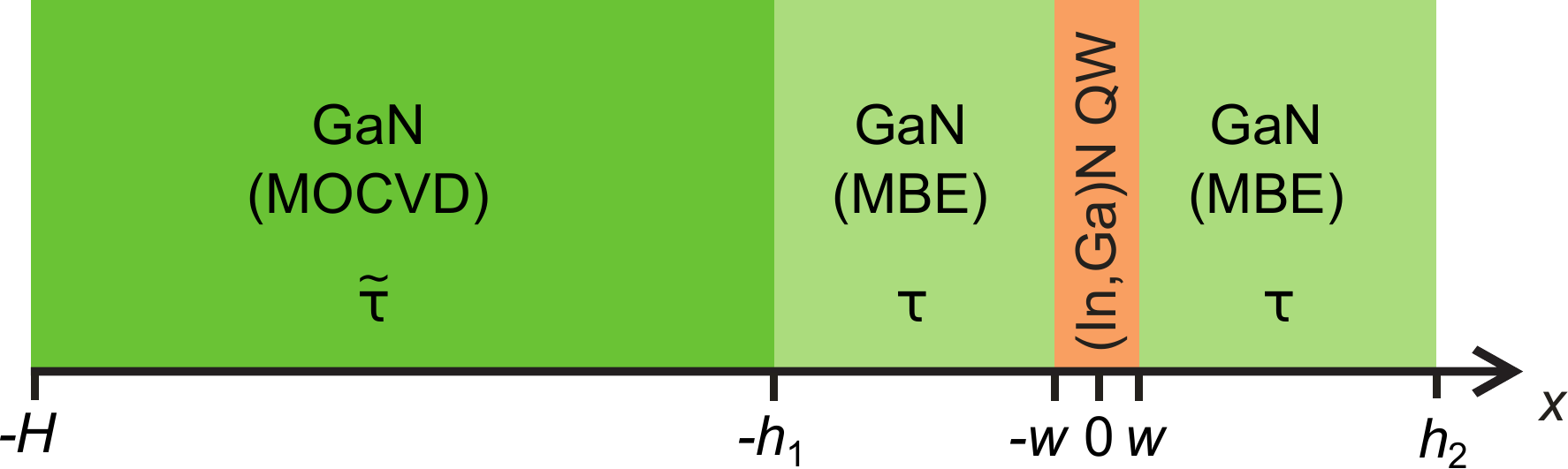} \caption{Geometry of the one-dimensional diffusion problem. The MBE-grown GaN layers on both sides of the (In,Ga)N quantum well and the MOCVD-grown GaN buffer layer possess different exciton lifetimes $\tau$ and $\tilde{\tau}$, respectively.}
\label{fig8} 
\end{figure}

Figure \ref{fig8} illustrates the geometry of the problem under consideration. The (In,Ga)N quantum well at $|x|<w$ is considered as a perfect sink for excitons that are created by the source on both sides of it. Quantum capture is taken into account in the most simple way possible, namely, by assuming an effective width of the quantum well larger than its actual one. Diffusion of excitons takes place in both the thick MOCVD-grown buffer layer ($-H<x<-h_{1}$) and the MBE-grown layers on top of the buffer below ($-h_{1}<x<-w$) and above ($w<x<h_{2}$) the well. The diffusion coefficient $D$ is assumed to be the same in both MBE-grown and MOCVD-grown GaN. For the GaN template fabricated by MOCVD, a larger fraction of excitons recombine radiatively, which is evident from the increasing intensity in the left parts of the GaN CL intensity profiles in Figs.~\ref{fig4} and \ref{fig5}. Hence, the effective lifetimes of excitons are different, denoted here by $\tilde{\tau}$ and $\tau$, which result in the respective quantum efficiencies $\tilde{\eta}=\tilde{\tau}/\tau_{r}$ and $\eta=\tau/\tau_{r}$ with the radiative lifetime $\tau_{r}$. 

The exciton density $n(x)$ is a solution of the one-dimensional diffusion equation 
\begin{equation}
D\frac{d^{2}n}{dx^{2}}-\frac{n}{\tau(x)}+\bar{Q}(x)=0,\label{eq:2}
\end{equation}
where $\tau(x)=\tilde{\tau}$ for $-H<x<-h_{1}$ and $\tau(x)=\tau$ for $x>-h_{1}$. The boundary condition in the buffer layer far from the well can be written as $n\left|_{x=-H}\right.=0$. Since the borders of the well $x=\pm w$ are perfect sinks for excitons, we have $n\left|_{x=\pm w}\right.=0$. The top GaN layer $w<x<h_{2}$ ends at the GaN(0001) facet. In the vicinity of this surface, the CL intensity is quenched strongly (see Fig.~\ref{fig5}]). This quenching is partly due to the fact that electrons are scattered out of the sample when the beam approaches the sample edge, thus reducing the number of cathodogenerated carriers, which we take into account in the simulations. However, to accurately reproduce the experimental profiles, we have to consider in addition substantial surface recombination taking place at this edge, possibly due to cleavage-induced defects. For simplicity, and since this boundary is far from the quantum well, we assume total exciton absorption at this boundary as expressed by the boundary condition $n\left|_{x=h_{2}}\right.=0$. The boundary problem with the piecewise constant function $\tau(x)$ is solved on the intervals of constant $\tau$ requiring the continuity of the solution $n$ and the flux $-D\,dn/dx$ at $x=-h_{1}$.

\subsubsection*{The Green function solution}

Since the diffusion equation (\ref{eq:2}) is linear, its solution can be represented as a convolution \begin{equation} n(x)=\intop_{-H}^{h_{2}}\bar{Q}(\xi)G(x,\xi)\,d\xi\label{eq:3} \end{equation} of the source $\bar{Q}(\xi)$ with the Green function $G(x,\xi)$, which is a solution of the diffusion equation for the unit point source at the point $\xi$, namely, 
\begin{equation}
D\frac{d^{2}G}{dx^{2}}-\frac{G}{\tau(x)}+\delta(x-\xi)=0,\label{eq:4}
\end{equation}
with the same boundary conditions as imposed on the exciton density $n(x)$.

The exciton flux to the well from both sides 
\begin{equation}
\mathcal{F}_{\mathrm{QW}}=-D\left.\frac{dn}{dx}\right|_{x=\pm w}\label{eq:5}
\end{equation}
is given by the flux to the well due to a unit source at $\xi$ 
\begin{equation}
F_{\mathrm{QW}}(\xi)=-D\left.\frac{dG(x,\xi)}{dx}\right|_{x=\pm w},\label{eq:6}
\end{equation}
convoluted with the source distribution, 
\begin{equation}
\mathcal{F}_{\mathrm{QW}}=\intop_{-H}^{h_{2}}\bar{Q}(\xi)F_{\mathrm{QW}}(\xi)\,d\xi.\label{eq:7}
\end{equation}

The light emission from the GaN matrix occurs with the quantum efficiency
$\eta(x)=\tau(x)/\tau_{r}$ of the respective layer. Hence, the GaN emission intensity is proportional to 
\begin{equation}
\mathcal{F}_{\mathrm{GaN}}=\intop_{-H}^{h_{2}}\eta(x)\frac{n(x)}{\tau(x)}\,dx=\frac{1}{\tau_{r}}\intop_{-H}^{h_{2}}n(x)\,dx.\label{eq:7a}
\end{equation}
Substituting Eq.~(\ref{eq:3}), we express the GaN intensity in the same form as Eq.~(\ref{eq:7}), where the intensity due to a point source at $\xi$ is 
\begin{equation}
F_{\mathrm{GaN}}(\xi)=\frac{1}{\tau_{r}}\intop_{-H}^{h_{2}}G(x,\xi)\,dx.\label{eq:7b}
\end{equation}

\subsubsection*{Flux determination based on a reciprocity relation}

The calculation of the fluxes can be simplified by solving an adjoint problem, where the source term is excluded from the diffusion equation and replaced by a modified boundary condition. This approach was used for the three-dimensional problem of the exciton diffusion at threading dislocations in GaN \cite{donolato_1985,sabelfeld_2017}. For the present problem, it can be explicitly formulated as follows.

Let us show that the function $F(x)$ in Eq.~(\ref{eq:6}) is the solution of the homogeneous diffusion equation 
\begin{equation}
D\frac{d^{2}F}{dx^{2}}-\frac{F}{\tau(x)}=0\label{eq:8}
\end{equation}
with the modified boundary condition $F\left|_{x=\pm w}\right.=1$ instead of $F\left|_{x=\pm w}\right.=0$. The other boundary condition remains unchanged, either $F\left|_{x=-H}\right.=0$ or $F\left|_{x=h_{2}}\right.=0$ for the fluxes to the well from different sides.

The proof is straightforward: we multiply Eq.~(\ref{eq:4}) with $F(x)$, Eq.~(\ref{eq:8}) with $G(x,\xi)$, subtract one from the other, and integrate over $x$. This yields 
\begin{equation}
D\intop\left(F\frac{d^{2}G}{dx^{2}}-G\frac{d^{2}F}{dx^{2}}\right)\,dx+F(\xi)=0.\label{eq:9}
\end{equation}
Since
\begin{equation}
F\frac{d^{2}G}{dx^{2}}-G\frac{d^{2}F}{dx^{2}}=\frac{d}{dx}\left(F\frac{dG}{dx}-G\frac{dF}{dx}\right), \label{eq:10}
\end{equation}
we can reduce the integral in Eq\@.~(\ref{eq:9}) to the values at the two ends of the respective integration interval. Since $G=0$ at both ends of the interval and $F\left|_{x=\pm w}\right.=1$ while $F\left|_{x=-H}\right.=0$ or $F\left|_{x=h_{2}}\right.=0$, Eq.~(\ref{eq:9}) reduces to Eq.~(\ref{eq:6}). Since the MOCVD-grown buffer layer is thick and the effect of its far end at $x=-H$ is negligible, we extend the buffer layer to infinity and take a boundary condition $F\left|_{x\rightarrow-\infty}\right.=0$. Then, we solve the homogeneous diffusion equation (\ref{eq:8}) and write the CL intensity of the well due to a point source at $x$ as 
\begin{equation}
F_{\mathrm{QW}}(x)=\begin{cases}
\exp\left[(x+h_{1})/\tilde{L}\right]/g(-w), & x<-h_{1},\\
g(x)/g(-w), & -h_{1}<x<-w,\\
1, & -w<x<w,\\
\dfrac{\sinh\left[(h_{2}-x)/L\right]}{\sinh\left[(h_{2}-w)/L\right]}, & w<x<h_{2},
\end{cases}\label{eq:11}
\end{equation}
where 
\begin{equation}
g(x)=\cosh[(h_{1}+x)/L]+(L/\tilde{L})\sinh[(h_{1}+x)/L]\label{eq:11a}
\end{equation}
and $L=\sqrt{D\tau}$, $\tilde{L}=\sqrt{D\tilde{\tau}}$ are the respective diffusion lengths.

To find the GaN intensity from a point source in the top layer, $w<x<h_{2}$, we calculate the fluxes to the two surfaces by solving the respective modified boundary problems. The remaining intensity, multiplied with the quantum efficiency $\eta$, is the GaN signal 
\begin{multline}
F_{\mathrm{GaN}}(x)=\eta\left(1-\dfrac{\sinh\left[(h_{2}-x)/L\right]+\sinh\left[(x-w)/L\right]}{\sinh\left[(h_{2}-w)/L\right]}\right)\\
(w<x<h_{2}).\label{eq:12}
\end{multline}

If the difference in quantum efficiencies between the MOCVD and the MBE layers is ignored, the GaN signal for $x<-w$ is calculated similarly, by subtracting the flux to the well: 
\begin{equation}
F_{\mathrm{GaN}}^{(0)}(x)=\eta\left(1-\exp\left[(x+w)/L\right]\right),\,\,\,\,\,\,\,(x<-w),\label{eq:13}
\end{equation}
where the superscript (0) indicates that the larger quantum efficiency of the MOCVD layer is ignored. The general case is presented below.

\subsubsection*{The adjoint problem for the GaN intensity}

Let us show now that the integral of the Green function for the source below the quantum well,
\begin{equation}
F_{\mathrm{GaN}}(\xi)=\frac{1}{\tau_{r}}\intop_{-H}^{-w}G(x,\xi)\,dx,\,\,\,\,\,-H<\xi<-w,\label{eq:14}
\end{equation}
is a solution of another adjoint problem, namely, the inhomogeneous equation 
\begin{equation}
D\frac{d^{2}F}{dx^{2}}-\frac{F}{\tau(x)}+\frac{1}{\tau_{r}}=0\label{eq:15}
\end{equation}
with unchanged boundary conditions $F\left|_{x=-w}\right.=0$ and $F\left|_{x=-H}\right.=0$. For a proof, we multiply the equation for the Green function (\ref{eq:4}) with $F$, the adjoint equation (\ref{eq:15}) with $G$, and subtract. Then, we get 
\begin{equation}
\frac{d}{dx}\left(F\frac{dG}{dx}-G\frac{dF}{dx}\right)=-F(x)\delta(x-\xi)+\frac{G(x)}{\tau_{r}}.\label{eq:16}
\end{equation}

Integrating this equation over the interval $-H<x<-w$, we reduce the left-hand part of the equation to the values at the ends of the interval. It is equal to zero, since both $F(x)$ and $G(x)$ are zero at both ends. The integral of the right-hand part gives Eq.~(\ref{eq:14}). Here, we consider the finite interval $-H<x<-w$ and do not extend the thickness of the buffer layer to infinity as done above in the calculation of the flux (\ref{eq:11}), since the boundary conditions would not be satisfied in this case.

To represent the solution of Eq.~(\ref{eq:15}) with boundary conditions $F\left|_{x=-w}\right.=0$ and $F\left|_{x=-H}\right.=0$ in a compact form, we introduce a function 
\begin{eqnarray}
P(x_{1},x_{2}) & = & (\tilde{L}/L)\cosh[(h_{1}+x_{1})/L]\sinh[(h_{1}+x_{2})/\tilde{L}]\nonumber \\
 &  & -\sinh[(h_{1}+x_{1})/L]\cosh[(h_{1}+x_{2})/\tilde{L}]\label{eq:17}
\end{eqnarray}
and denote $P_{0}=-P(-w,-H)$. Then, the solution is
\begin{widetext}
\begin{equation}
F_{\mathrm{GaN}}(x)=\begin{cases}
\cfrac{1}{P_{0}}\left[\left(\tilde{\eta}+(\eta-\tilde{\eta})\cosh\cfrac{H-h_{1}}{\tilde{L}}\right)P(-w,x)-\eta\cfrac{\tilde{L}}{L}\sinh\cfrac{H+x}{\tilde{L}}\right]+(\eta-\tilde{\eta})\cosh\cfrac{h_{1}+x}{\tilde{L}}+\tilde{\eta}, & -H<x<-h_{1},\\
\cfrac{1}{P_{0}}\left[\eta P(x,-H)+(\eta-\tilde{\eta})\sinh\cfrac{x+w}{L}\cosh\cfrac{H-h_{1}}{\tilde{L}}+\tilde{\eta}\sinh\cfrac{x+w}{L}\right]+\eta, & -h_{1}<x<-w.
\end{cases}\label{eq:18}
\end{equation}
\end{widetext}

This solution, together with Eq.~(\ref{eq:12}), provides an expression for the GaN intensity for any position $\xi$ of the point source. In the limit of equal effective lifetimes in the layers, $\tilde{\tau}=\tau$ (and hence equal quantum efficiencies and diffusion lengths, $\tilde{\eta}=\eta$ and $\tilde{L}=L$), Eq.~(\ref{eq:18}) becomes 
\begin{equation}
F_{\mathrm{GaN}}(x)=\eta\left[1-\frac{\sinh[(H+x)/L]-\sinh[(x+w)/L]}{\sinh[(H-w)/L]}\right],\label{eq:19}
\end{equation}
similar to Eq.~(\ref{eq:12}). It reduces to Eq.~(\ref{eq:13}) in the limit $H\rightarrow\infty$.

\subsubsection*{Application to the case studied by \citet{netzel_2020}} 

The same formalism can be applied to obtain an exact analytical solution of the exciton diffusion problem for the experiment by \citet{netzel_2020}. In this experiment, two different samples consisting of (In,Ga)N/GaN QWs buried at different depths $d_1$ and $d_2 > d_1$ from the top GaN(0001) surface are excited by diffusion of carriers photogenerated in the cap layer in a depth $\alpha^{-1} < d_1$, where $\alpha$ is the absorption coefficient  of GaN. A diffusion length $L_{\text{N}}$ is derived from the resulting experimental photoluminescence intensities $I_{\text{exp}}(d_{1})$ and $I_{\text{exp}}(d_{2})$, assuming that it is related to these intensities by the simple expression
\begin{equation}
\frac{I_{\text{exp}}(d_{2})}{I_{\text{exp}}(d_{1})}=\exp\left(-\frac{d_{2}-d_{1}}{L_{\text{N}}}\right).\label{eq:20}
%L_{\mathrm{N}}=\frac{d_{2}-d_{1}}{\ln[I_{\text{exp}}(d_{1})/I_{\text{exp}}(d_{2})]}.\label{eq20}
\end{equation}

\begin{figure}
\includegraphics[width=0.85\columnwidth]{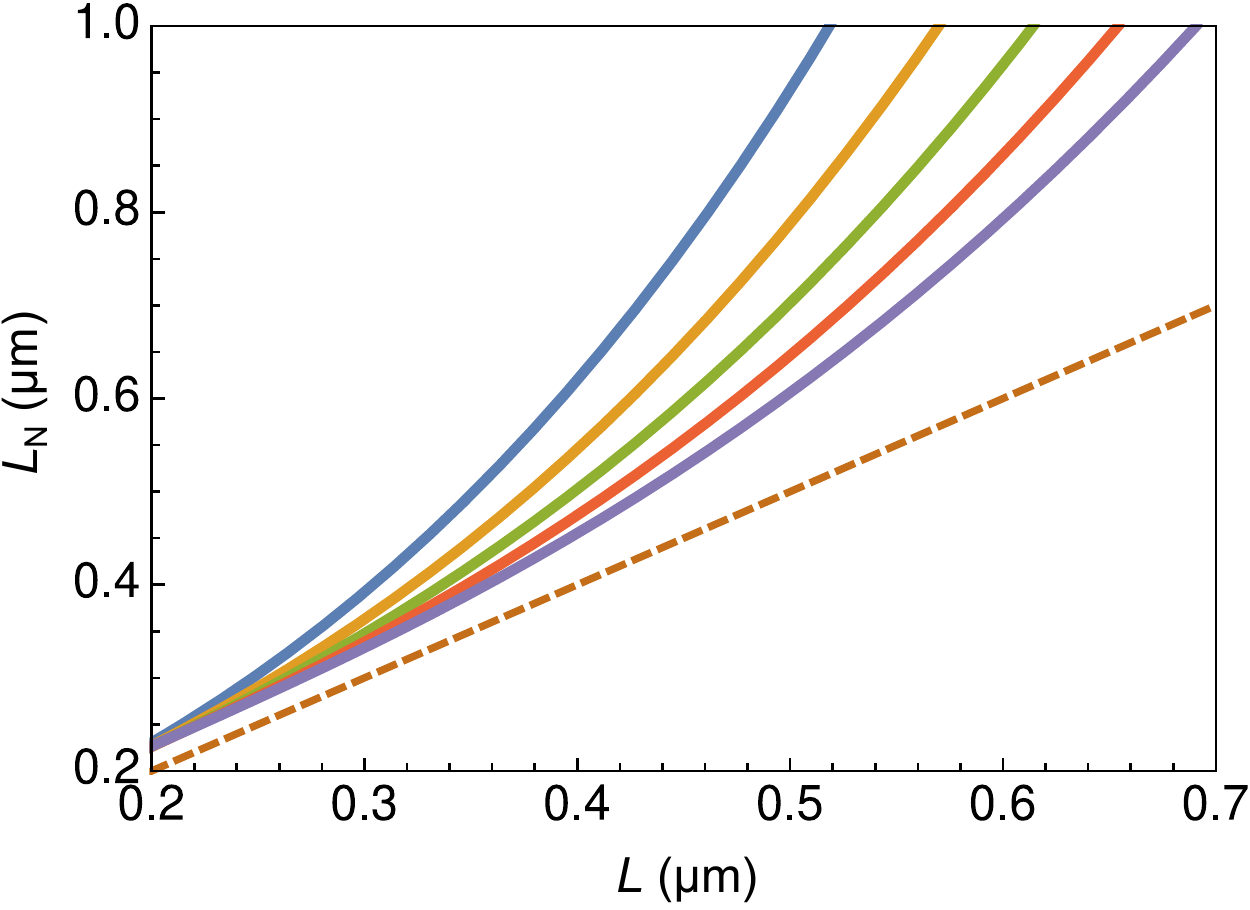} 
\caption{Dependence of the apparent diffusion length $L_\text{N}$ determined in Ref.~\onlinecite{netzel_2020} on the actual diffusion length $L$. The dashed line corresponds to $L_\text{N} = L$. The solid lines represent the values of $L_\text{N}$ for a surface recombination velocity $S$ ranging, from left to right, from zero to $8 \times 10^4$~cm/s in steps of $2 \times 10^4$~cm/s. All other parameters for the calculation are taken from Ref.~\onlinecite{netzel_2020}: $\alpha = 1.2 \times 10^5$~cm$^{-1}$, $d_1 = 0.3$~\textmu m, $d_2 = 0.45$~\textmu m. The diffusivity $D$ is substituted with $L^2/\tau$, and the carrier/exciton lifetime $\tau$ is taken to be 500~ps, being close to the values measured in Ref.~\onlinecite{netzel_2020} at both low and high temperatures.}
\label{fig9} 
\end{figure}

Using the same Green function formalism as above, we can represent the flux to the QW in the same way as in Eq.~(\ref{eq:7}),
\begin{equation}
I(d)=\intop_{0}^{d}\bar{Q}(x)F(x)\,dx.\label{eq:21}
\end{equation}
Here, $\bar{Q}(x)=\exp(-\alpha x)$ is the density of photoexcited carriers, and $F(x)$ is the flux to the QW from a unit source at depth $x$. This flux is a solution of the homogeneous diffusion equation (\ref{eq:8}) with the modified boundary condition at the QW: instead of the boundary condition for the carrier density $n|_{x=d}=0$, we write $F|_{x=d}=1$. The boundary condition at the top surface $x=0$ remains unchanged. Taking into account the finite surface recombination velocity $S$ at the GaN(0001) surface, the boundary condition thus reads 
\begin{equation}
\left.\left(-D\frac{dF}{dx}+SF\right)\right|_{x=0}=0.\label{eq:22}
\end{equation}
The solution of this boundary problem is
\begin{equation}
F(x)=\frac{\exp(x/L)+\beta\exp(-x/L)}{\exp(d/L)+\beta\exp(-d/L)},\label{eq:23}
\end{equation}
where $\beta=(D-LS)/(D+LS)$. Calculation of the integral (\ref{eq:21}) with the source $\bar{Q}(x)=\exp(-\alpha x)$ and the function $F(x)$ given by Eq.~(\ref{eq:23}) is straightforward, and the result is not presented here explicitly.  With the intensities $I(d)$ thus calculated for two QW depths $d_{1}$ and $d_{2}$, we can find the dependence of the length $L_{\mathrm{N}}$ on the actual diffusion length $L$ as in Eq.~(\ref{eq:20}), but with the intensities from Eq.~(\ref{eq:21}):
\begin{equation}
L_{\mathrm{N}}=\frac{d_{2}-d_{1}}{\ln[I(d_{1})/I(d_{2})]}.\label{eq:24}
\end{equation}

Figure \ref{fig9} shows the results of Eq.~(\ref{eq:24}) for values of the surface recombination velocity $S$ from zero to $8 \times 10^4$~cm/s, covering the range reported for GaN(0001) surfaces (mid-$10^3$ to mid-$10^4$~cm/s) \cite{onuma_2012,bulashevich_2016}. Evidently, the deviation of $L_\text{N}$ from $L$ increases drastically for $L > d_1$ and with decreasing $S$.  For $S = 4 \times 10^4$~cm/s, Eq.~(\ref{eq:24}) overestimates the actual diffusion length by 15\% for high temperatures, where $L_\text{N} \approx L$, and by as much as 40\% at low temperatures, at which $L > d_2$.  

\section{Determination of the effective lifetime from photoluminescence transients} 
\label{appendix:lifetime}

\begin{figure}
\includegraphics[width=0.9\columnwidth]{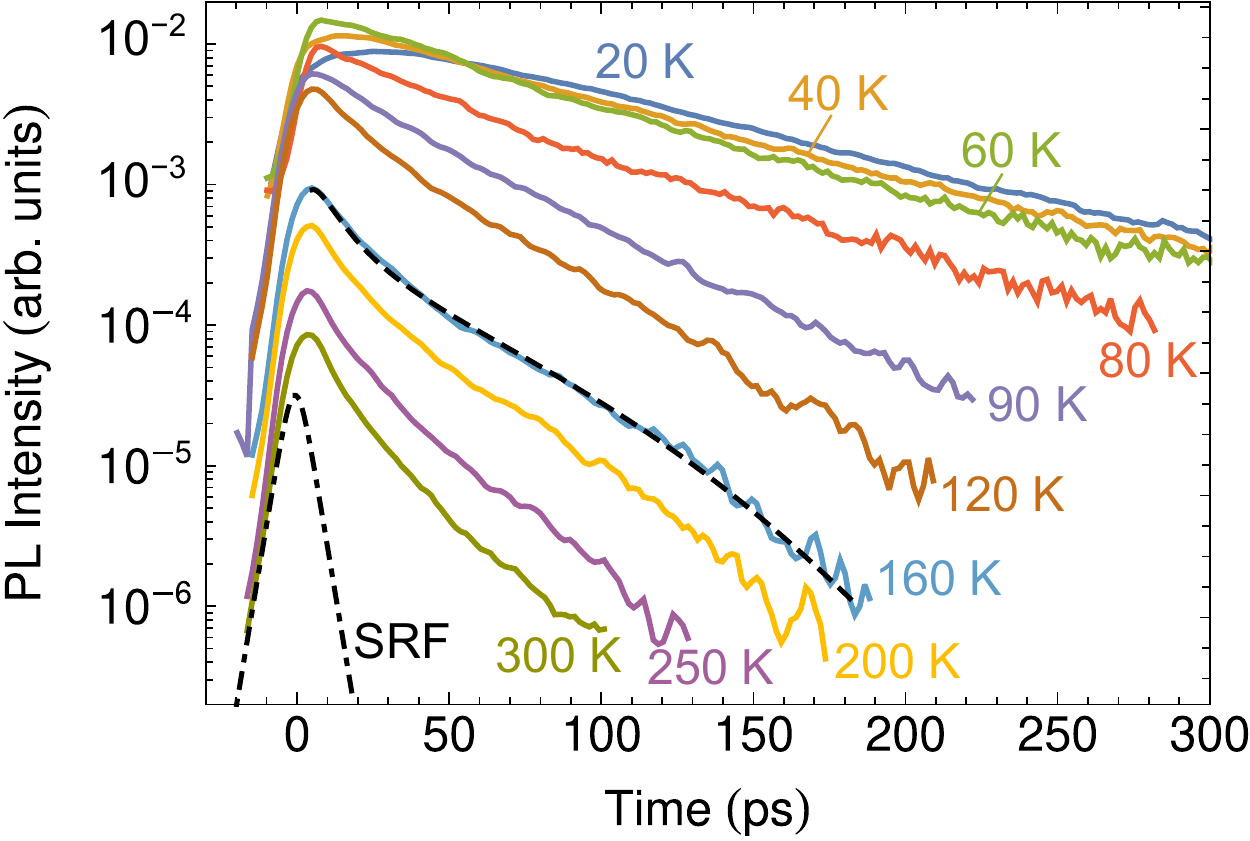} 
\caption{Photoluminescence intensity transients obtained by spectrally integrating over the exciton line in streak camera images acquired at temperatures between 20 and 300~K. The dashed line represents an example for a transient governed by Shockley-Read-Hall recombination, while the dash-dotted line shows the system response function (SRF).}
\label{fig10} 
\end{figure}

Figure \ref{fig10} shows exemplary photoluminescence intensity transients acquired from the top GaN(0001) surface of the sample under investigation. At low temperatures, the decay is strictly monoexponential and is thus characterized by a single effective lifetime $\tau$. This time is too short ($< 100$~ps) to be governed by radiative processes. For temperatures higher than 60~K, the decay decreases in overall intensity, becomes progressively faster, and acquires an increasingly nonexponential shape with a fast initial decay. The former two of these observations reflect the fact that carrier recombination is governed by nonradiative channels. The latter one may have various different reasons. For example, laser stray light may result in a fast initial component, with an apparent decay time reflecting the instrumental resolution. The system response function (SRF) for the present experiments is also shown in Fig.~\ref{fig10}, and can be represented by a hyperbolic secant $\text{sech}\left( t/\Delta t \right)$ with $\Delta t = 3.3$~ps. Clearly, the temporal system response is significantly faster than the initial decay component, ruling out this instrumental artifact as its origin. Second, a mechanism known to result in nonexponential transients is surface recombination \cite{*[{}] [{. Note that Eq.~(167) on p.~105 contains a misprint: the term $\exp(S^2/D)$ should read $\exp(S^2 t/D)$.}] Ahrenkiel_1993}. The strength of surface recombination (the surface recombination velocity) is characteristic for a given surface, and the associated fast initial decay at higher temperatures would thus be expected to be a universal feature in the photoluminescence transients of GaN(0001) films. However, this fast decay would correspond to surface recombination velocities exceeding $10^6$~cm/s, clearly inconsistent with the range of values (mid-$10^3$ to mid-$10^4$~cm/s) observed for GaN(0001) films \cite{onuma_2012,bulashevich_2016}. Hence, we can also exclude surface recombination as being responsible for the fast initial component observed for the sample under investigation. 

These considerations imply that the shape of the transient is determined by nonradiative processes taking place in the bulk, thus being of relevance for the diffusion length. The most simple scenario consistent with the shape of the transients is electron-hole recombination via a single Shockley-Read-Hall center with strongly different electron and hole capture coefficients, which thus has the character of a trap \cite{Brandt_1996}. An exemplary transient based on this mechanism is compared in Fig.~\ref{fig10} to the experimental transient obtained at 160~K. Note that this model, while helping to understand the origin of the decay, does not characterize the decay in terms of an effective lifetime as required for determining the carrier diffusivity. To resolve this difficulty, we note that a lifetime $\tau$ can be extracted for any given transient, regardless of its shape, by simply  integrating over the normalized transient. The physical meaning of $\tau$ can be understood by considering that the area under the transient is proportional to the  internal quantum efficiency $\eta = \tau/\tau_r$, while the peak intensity is proportional to the radiative rate $1/\tau_r$ \cite{Brandt_1996}. Since the proportionality constant is the same, the generally unknown radiative lifetime drops out. Hence, integrating the normalized transients yields an effective lifetime equal to the actual one for a monoexponential decay, and an average effective lifetime proportional to the internal quantum efficiency for arbitrary decay shapes.     

\section{Simulation of carrier and exciton diffusivity} 
\label{appendix:mobility}
 
We obtain the diffusivity $D$ for free carriers and excitons using the established expressions for the drift mobility $\mu$ of particles with a density-of-states-mass $m_\text{p}$. The expressions used are valid only for nondegenerate semiconductors and are given below for the five major scattering mechanisms considered in this work.

$D$ is related to $\mu$ via the Einstein relation  
\begin{equation}
	D=\frac{k_B T \mu }{e_0}
\end{equation}
with the Boltzmann constant $k_B$ and the elementary charge $e_0$, and $\mu$ is being considered to result from the addition of independent scattering rates (Matthiessen's rule):

\begin{equation}
		\mu = \left(\sum_i \frac{1}{\mu_{\text{i}}}\right)^{-1}.
\end{equation}

The most important parameter in the expressions for $\mu$, regardless of the scattering mechanism, is the particle mass $m_p$. For semiconductors with anisotropic and nonparabolic valence bands such as GaN, the mass depends on the wave vector $\boldsymbol{k}$ and the energy $E$, and the quantity entering the expressions for the mobility is an appropriately defined density-of-states mass, which, in principle, depends on temperature \cite{wellenhofer_1997}. As an approximation of this temperature-dependent mass, two different sets of valence band dispersion parameters are usually derived, one yielding the hole masses directly at $\boldsymbol{k} = 0$ (strictly valid only for $T \rightarrow 0$), the other one also for finite $\boldsymbol{k}$ values away from the $\Gamma$ point (important at, for example, room temperature) \cite{chuang_1996,rinke_2008}. Free holes are dominant at high temperatures, and thus occupy states with $\boldsymbol{k}>0$ such that the latter of these valence band dispersion parameters applies. In contrast, excitons are dominant at low temperatures, and thus predominantly experience the valence band dispersion at $\boldsymbol{k} = 0$ \cite{hagele_1999}. Using the band dispersion parameters reported in Refs.~\onlinecite{rinke_2008} and \onlinecite{punya_2012}, these considerations result in density-of-states masses of $m_e = 0.2$, $m_h = 1.9$, and $m_x=0.8$, for electrons, holes, and excitons, respectively. Due to the temperature-dependent coexistence of free carriers and excitons [cf.\ Eq.~\ref{eq:saha} and Fig.~\ref{fig7}(a)], these values lead to a smooth but almost step-like variation of the particle mass with temperature, comparable to the behavior of the thermal density-of-states mass derived in Ref.~\onlinecite{wellenhofer_1997} for SiC, which features an electronic band structure very similar to GaN.

\subsubsection*{Neutral impurity scattering}

This mechanism is most important for excitons at low temperatures with $N_n$ being the density of impurities acting as neutral scattering centers \cite{seeger_1989}:

\begin{equation}
	\mu _{ni}=\frac{e_0^3 m\!_{p}}{80 \pi \epsilon_s \epsilon_0 \hbar^3 N_n},
	\label{eq:ni}
\end{equation}
where $\hbar$ is the reduced Planck constant. The material constant $\epsilon_s = 9.8$ is the static relative permitivity of GaN.

\subsubsection*{Ionized impurity scattering}

This mechanism is the dominant one for free carriers at low temperatures with $N_i$ being the density of impurities acting as ionized scattering centers. The situation we are dealing with is slightly different from the one encountered usually. First of all, we are interested in the diffusivity of both majority and minority carriers, with the latter being the one determining the ambipolar diffusivity. Second, the free carrier and the ionized impurity concentration decouple at low temperatures, since the background electron density rapidly decreases with decreasing temperature, and the total free carrier concentration is determined by the cathodogenerated one. In general, $n = n_0 + \Delta n$, $N_D^+ = n_0 + N_A^-$, and $N_i = n_0 + 2 N_A$ [see Eq.~(\ref{eq:carrconc})]. For $T \rightarrow 0$, $n_0 \rightarrow 0 $, and the scenario becomes entirely symmetric, such that a distinction between majority and minority carriers is no longer meaningful. This situation cannot be treated adequately by either the Conwell-Weisskopf or the Brooks-Herring approximations \cite{seeger_1989,chattopadhyay_1981} in the full temperature range, but requires the use of the generalized model developed by \citet{ridley_1977}: 

\begin{equation} 
	\mu _{ii}=\frac{2^{15/2} \sqrt{\pi} \left(\epsilon_s \epsilon_0\right)^2 \left(k_B T\right)^{3/2}}{\sqrt{m\!_{p}} e_0^3 N_i \mathcal{L}},
	\label{eq:ii}
\end{equation}
where $\mathcal{L}$ is the equivalent of the logarithmic terms in the Conwell-Weisskopf and the Brooks-Herring approximations:

\begin{align}
	\mathcal{L} & = \exp\left(\frac{\alpha \beta}{1+\beta}\right) \left\{ E_1\left(\frac{\alpha \beta}{1+\beta}\right)- E_1\left(\alpha \beta \right)\right. \\ \nonumber
	& - \left.\frac{1}{\alpha\beta}\left[\exp\left(-\frac{\alpha \beta}{1+\beta}\right) - \exp\left(-\alpha \beta\right)\right]\right\} 
\end{align}

with the exponential integral $E_1(z)$,

\begin{equation}
	\alpha = \frac{e_0^4 N_i^{2/3}}{576 \pi (\epsilon_s \epsilon_0)^2 (k_B T)^2},
\end{equation}

and

\begin{equation}
	\beta = \frac{24 m\!_{p} \epsilon_s \epsilon_0 (k_B T)^2}{e_0^2 \hbar^2 n}.
\end{equation}

\subsubsection*{Piezoelectric scattering}

For any piezoelectric semiconductor, this mechanism is important for free carriers at low to intermediate temperatures \cite{seeger_1989,shur_1996}:

\begin{equation}
	\mu _{pz}=\frac{2^{9/2} \sqrt{\pi } \hbar ^2 \epsilon_s \epsilon_0}{3 m\!_{p}^{3/2} e_0 K_{\text{av}}^2
   \sqrt{k_B T}},
   	\label{eq:piezo}
\end{equation}
with the electromechanical couping coefficient $K_{\text{av}}$. For free carriers, we assume $K_{\text{av}} =0.093 $ (cf.\ Refs.~\onlinecite{look_2001,vitanov_2007}). The coupling of excitons to the piezoelectric field of acoustic vibrations is expected to be weaker \cite{zinovev_1983}, but since solid data are lacking, we treat $K_{\text{av}}$ as a free parameter for excitons. 

\subsubsection*{Deformation potential scattering}

For any semiconductor, this mechanism is important for both free carriers and excitons at intermediate temperatures \cite{seeger_1989,shur_1996}.

\begin{equation}
	\mu _{ac}=\frac{2^{3/2} \sqrt{\pi }\hbar ^4 c_l}{3 m\!_{p}^{5/2} e_0 D_{ac,p}^2 \left(k_B T\right)^{3/2}}.
	\label{eq:def}
\end{equation}
The acoustic deformation potential is taken as $D_{ac,e} = 8.3$~eV \cite{shur_1996}, $D_{ac,h} = -9.1$~eV \cite{wiley_1970,oguzman_1996}, and $D_{ac,x} = E_{1,e}-E_{1,h}$ \cite{selbmann_1996,oki_2017}.

\subsubsection*{Polar optical scattering}

At elevated temperatures, the diffusivity of both free carriers and excitons is limited by polar optical scattering \cite{shur_1996}:

\begin{equation}
	\mu _{po}=\frac{4 \pi  \epsilon _0 \kappa  \hbar ^2 \left(1-5 k_B T/E_G\right)}{e_0 m\!_{p}^{3/2} \sqrt{2 E_{\text{LO}} \left(1+E_{\text{LO}}/E_G\right)}N_{\text{LO}}},
	\label{eq:polar}
\end{equation}
 with the longitudinal optical (LO) phonon energy of $E_{\text{LO}}=92$~meV, the bandgap $E_{G}$, the phonon occupation number $N_{\text {LO}} = \left[\exp(E_{LO}/k_B T)-1\right]^{-1}$, and the LO coupling constant of $\kappa = \left(1/\epsilon_s + 1/\epsilon_{\infty}\right)^{-1}$ with the high-frequency relative permitivity $\epsilon_\infty = 5.35$ of GaN.

\bibliography{references}
\end{document}